
\def\pmb#1{\setbox0=\hbox{$#1$}%
\kern-.025em\copy0\kern-\wd0
\kern.05em\copy0\kern-\wd0
\kern-.025em\raise.0433em\box0 }
\def\diag{\,{\rm diag}\,}
\def\square{\kern1pt\vbox{\hrule height 1.2pt\hbox{\vrule width 1.2pt\hskip 3pt
   \vbox{\vskip 6pt}\hskip 3pt\vrule width 0.6pt}\hrule height 0.6pt}\kern1pt}
\def\upleftarrow#1{\raise2ex\hbox{$\leftarrow$}\mkern-16.5mu #1}
\def\uprightarrow#1{\raise2ex\hbox{$\rightarrow$}\mkern-16.5mu #1}
\def\upleftrightarrow#1{\raise1.5ex\hbox{$\leftrightarrow$}\mkern-16.5mu #1}
\def\ll{\left\langle}
\def\rr{\right\rangle}
\magnification=1200
\vsize=7.5in
\hsize=5.6in
\tolerance 10000
\def\equalsone{1\hskip-.29em1}

\baselineskip 12pt plus 1pt minus 1pt
\centerline{\bf POINCAR\'E GAUGE THEORY FOR GRAVITATIONAL}
\smallskip
\centerline{{\bf FORCES IN $\pmb{(1+1)}$ DIMENSIONS
}\footnote{*}{This
work is supported in part by funds
provided by the U. S. Department of Energy (D.O.E.) under contract
\#DE-AC02-76ER03069, and by the Swiss National Science Foundation.}}
\vskip 24pt
\centerline{D.~Cangemi and R.~Jackiw}
\vskip 12pt
\centerline{\it Center for Theoretical Physics}
\centerline{\it Laboratory for Nuclear Science}
\centerline{\it and Department of Physics}
\centerline{\it Massachusetts Institute of Technology}
\centerline{\it Cambridge, Massachusetts\ \ 02139\ \ \ U.S.A.}
\vskip 1.5in
\centerline{Submitted to: {\it Annals of Physics\/}}
\vfill
\centerline{ Typeset in $\TeX$ by Roger L. Gilson}
\vskip -12pt
\noindent CTP\#2165\hfill February 1993
\eject
\baselineskip 24pt plus 2pt minus 2pt
\centerline{\bf ABSTRACT}
\medskip
We discuss in detail how string-inspired lineal gravity can be formulated as
a gauge theory based on the centrally extended Poincar\'e group in
$(1+1)$ dimensions.  Matter couplings are constructed in a gauge invariant
fashion, both for point particles and Fermi fields.  A covariant tensor
notation is developed in which gauge invariance of the formalism is
manifest.
\vfill
\eject
\noindent{\bf I.\quad INTRODUCTION}
\medskip
\nobreak
In this article we elaborate on our two recent Letters,$^{1,\,2}$ which deal
with $(1+1)$-dimensional gravity theory.  In the first of these, we gave a
gauge theoretical formulation for the ``string-inspired'' model of ``dilaton''
gravity,$^{3,\,4}$ complementing an analogous discussions based on the
de~Sitter group for the constant curvature model$^5$ that had been introduced
earlier.  The gauge group for the ``string-inspired'' model is the Poincar\'e
group,$^{4,\,6}$
but surprisingly a central extension in the algebra is needed for
a manifestly invariant description of the cosmological constant.$^{1,\,7}$
The second Letter was devoted to
a gravity-matter interaction that modifies in a fashion specific to $(1+1)$
dimensions the usual geodesic equation of motion for point particles
without spoiling general covariance.  Also we showed that this additional
interaction fits neatly into the gauge theoretical formulation based on the
extended Poincar\'e group, the extension being responsible for the new
interaction.

Here we give a fuller account of these matters.  We begin in Section~II by
determining how the geodesic equation can be altered while still preserving
general covariance.  Possible  additions describe novel gravitational forces
on point particles and novel gravitational interactions for fields.  We
examine in detail the modifications for point particles and for
 Fermi fields.  We also compute the functional determinant for massless
fermions interacting with gravity, when the
conventional interaction is supplemented by our additions.

The above development is presented in the geometric formalism, using the
metric tensor and/or the {\it Zweibein\/}.  We then rederive the
equations in a gauge theoretical formalism based on the extended Poincar\'e
group.  To this end, general properties of the extended Poincar\'e group are
summarized in Section~III, while Section~IV is a brief reprise of Section~II,
but now in group theoretical language, and it is shown that the novel
interactions have a natural setting here. Our
approach to the construction of matter interactions that are
Poincar\'e gauge invariant
follows Grignani and Nardelli,$^7$ but we also present a manifestly
covariant tensor formalism.

The discussion of matter dynamics in Sections~II and IV is carried out without
specifying the gravitational action.  In Section~V, the gravity sector of the
theory is described in terms of connections for the extended Poincar\'e gauge
group.  Concluding remarks comprise the last Section~VI.

Our conventions are the following.  Velocity of light is scaled to unity.
Lower case Greek and Latin letters refer respectively to space-time and
tangent space components.  The former are raised and lowered by the metric
tensor $g_{\mu\nu}$; the latter are moved by the Minkowski-space metric $h_{ab}
=
\diag (1,-1)$.  The {\it Zweibein\/} $e^a_\mu$ is related to $g_{\mu\nu}$ by
$$g_{\mu\nu} = e^a_\mu e^b_\nu h_{ab}\eqno(1.1)$$
Also we use the anti-symmetric tangent space tensor $\epsilon^{ab}$,
normalized by $\epsilon^{01} = 1=-\epsilon_{01}$,
as well as the numerical tensor densities
$\epsilon^{\mu\nu}$ and $\epsilon_{\mu\nu}$,
where $\epsilon^{\mu\nu}/\sqrt{-g}$ is a contravariant
anti-symmetric tensor and $\sqrt{-g}\,\epsilon_{\mu\nu}$ is covariant, with
$$\eqalign{g &\equiv \det g_{\mu\nu}\cr
\sqrt{-g} &= \det e^a_\mu = - {1\over 2} e^a_\mu e^b_\nu \epsilon^{\mu\nu}
\epsilon_{ab}\cr}\eqno(1.2)$$
The inverse $E^\mu_a$ of the {\it Zweibein\/} is given by
$$E^\mu_a = -{1\over\sqrt{-g}} \epsilon^{\mu\nu} \epsilon_{ab} e^b_\nu
\eqno(1.3)$$
We shall need the spin-connection $\omega_\mu$, either as an independent
variable or determined by the {\it Zweibein\/}.
$$\omega_\mu = {1\over\sqrt{-g}}
\epsilon^{\alpha\beta}e^a_\mu h_{ab}  \partial_\alpha
e^b_\beta\eqno(1.4)$$
This follows from the torsion-free condition.
$$\epsilon^{\mu\nu} \left( \partial_\mu e^a_\nu + \epsilon^a{}_b \omega_\mu
e^b{}_\nu\right) = 0 \eqno(1.5)$$
We introduce the Christoffel connection $\Gamma^\alpha_{\mu\nu}$ by
$$\partial_\mu e^a_\nu + \epsilon^a{}_b \omega_\mu e^b{}_\nu -
\Gamma^\alpha_{\mu\nu} e^a_\alpha = 0 \eqno(1.6)$$
and Eq.~(1.5) insures that $\Gamma^\alpha_{\mu\nu}$ is given by the usual
formula.
$$\Gamma^\alpha_{\mu\nu} = {1\over 2} g^{\alpha\beta} \left( \partial_\mu
g_{\beta\nu} + \partial_\nu g_{\beta\mu} - \partial_\beta
g_{\mu\nu}\right)\eqno(1.7)$$
The scalar curvature $R$ is constructed from the spin connection.
$$\partial_\mu \omega_\nu - \partial_\nu \omega_\mu =-{1\over 2}\sqrt{-g}\,
\epsilon_{\mu\nu} R \eqno(1.8)$$

When dealing with a two-component
Fermi field $\psi$, we use $2\times 2$ Dirac matrices
$\gamma_a$ satisfying
$$\left\{ \gamma_a,\gamma_b\right\} = 2h_{ab}\eqno(1.9)$$
Also defined is the Dirac--Hermitian chiral matrix $\gamma_5$,
$$\gamma_5 = i\gamma_0\gamma_1\ \ ,\qquad \gamma^\dagger_5=-\gamma_5
\eqno(1.10)$$
which dualizes the gamma matrices.
$$\gamma_a\gamma_5 = i\epsilon_{ab}\gamma^b\eqno(1.11)$$

A tangent space Lorentz transformation,
$$\Lambda^a{}_b \equiv \delta^a{}_b \cosh\alpha + \epsilon^a{}_b
\sinh\alpha\eqno(1.12)$$
where the rapidity $\alpha$ can be a function of the space-time point $x^\mu$
when the transformation is local, acts on tangent space indices.  Explicitly,
$$\eqalignno{e^a_\mu &\longrightarrow \left( \Lambda^{-1}\right)^a{}_b e^b_\mu
&(1.13)\cr\noalign{\hbox{the spin connection transforms as a gauge potential,}}
\omega_\mu &\longrightarrow \omega_\mu + \partial_\mu\alpha
&(1.14)\cr\noalign{\hbox{and a Fermi field transforms by}}
\psi&\longrightarrow e^{{i\over 2}\alpha \gamma_5}\psi
&(1.15)\cr\noalign{\hbox{so
that a derivative supplemented by the spin connection transforms
covariantly.}}
\left( \partial_\mu - {i\over 2}\omega_\mu \gamma_5\right)\psi
&\longrightarrow e^{{i\over 2}\alpha\gamma_5} \left( \partial_\mu - {i\over
2}\omega_\mu\gamma_5\right) \psi &(1.16)\cr}$$
\goodbreak
\bigskip
\noindent{\bf II.\quad MATTER-GRAVITY INTERACTIONS}
\medskip
\nobreak
\noindent{\bf A.\quad Point Particle}
\medskip
\nobreak
The usual action for a material point particle of mass $m$, moving on the
world line $x^\mu(\tau)$, is constructed from the arc length.
$$I_m = - m \int ds = - m \int d\tau \sqrt{\dot x^\mu(\tau) g_{\mu\nu} \bigl(
x(\tau)\bigr) \dot x^\nu(\tau)}\eqno(2.\hbox{A}1)$$
The overdot denotes differentiation with respect to $\tau$, which parametrizes
the world line in an arbitrary way --- $I_m$ is parametrization invariant.
However, the $(1+1)$-dimensional setting provides additional, dimension
specific possibilities for matter-gravity interactions.

Let us observe that a force ${\cal F}^\mu$ added  to the geodesic
equation of motion
$$\eqalign{
g^{\mu\nu} \bigl( x(\tau)\bigr) {\delta\over\delta x^\nu (\tau)} I_m &= {d\over
d\tau}\ {1\over N} \dot x^\mu + {1\over N} \dot x^{\alpha}
\Gamma^\mu_{\alpha\beta} \dot x^{\beta} = {\cal F}^\mu\cr
N &\equiv {1\over m} \sqrt{\dot x^\alpha g_{\alpha\beta}\dot x^\beta} \cr}
\eqno(2.\hbox{A}2)$$
must satisfy various consistency conditions. First, to maintain
parametrization
invariance, ${\cal F}^\mu$ must be linear in $\dot x^\mu$.  Also the
transversality of the left side in (2.A2) to $\dot x^\mu$
enforces that condition on ${\cal
F}_\mu$.
$$\dot x^\mu {\cal F}_\mu = 0 \eqno(2.\hbox{A}3)$$
An option we do {\it not\/} take is ${\cal F}^\mu$ proportional to $\dot
x^\mu$, for then (2.A3) could only be true for massless particles, and also the
force would be dissipative. We are thus led to the formula
$${\cal F}_\mu = {\cal F}_{\mu\nu} \dot x^\nu\eqno(2.\hbox{A}4)$$
where ${\cal F}_{\mu\nu}$ is anti-symmetric.  It is clear that an {\it
externally\/} prescribed ${\cal F}_{\mu\nu}$ will result in loss of general
covariance. In order to maintain that principle, ${\cal F}_{\mu\nu}$ must be
constructed from the dynamical variables of the theory, and experience with
electromagnetism (which is {\it not\/} included in the above discussion) shows
that general covariance is preserved when ${\cal F}_{\mu\nu}$ is a
second-rank, anti-symmetric tensor.  In dimensions greater than two, such a
tensor cannot be constructed from particle and/or gravitational variables; it
arises when electromagnetic (or other gauge-field) degrees of freedom are
dynamically active --- but here we do not include such additional variables.
However, in two dimensions, gravitational variables allow constructing the
required tensor,
$${\cal F}_{\mu\nu} = \sqrt{-g}\,\epsilon_{\mu\nu} {\cal F}
\eqno(2.\hbox{A}5\hbox{a})$$
where ${\cal F}$ is a scalar, which evidently can be  a constant
or built from the scalar curvature.  We choose
the two simplest contributions to ${\cal F}$, which will be seen to fit very
naturally into our gauge theoretical formulation.
$${\cal F}= - {{\cal A}\over 2} R - {\cal B} \eqno(2.\hbox{A}5\hbox{b})$$
Here ${\cal A}$ and ${\cal B}$
are constants setting the strength of the addition.
The covariant, two-dimensional equation of motion, involving matter and
metric variables and generalizing the usual geodesic equation reads
$${d\over d\tau}\ {1\over N} \dot x^\mu + {1\over N} \dot x^\alpha
\Gamma^\mu_{\alpha\beta} \dot x^\beta + \left({1\over 2}{\cal A}R+{\cal B}
\right) g^{\mu\alpha}\sqrt{-g}\, \epsilon_{\alpha\beta}\dot x^\beta = 0
\eqno(2.\hbox{A}6)$$

The first addition, involving ${\cal A}R$, is
non-minimal, vanishing in the absence of curvature. It will be seen that this
term produces interaction with curvature familiar from conformal improvements
of dynamics.
The second addition$^2$ to the interaction (2.\hbox{A}6) is similar to
a constant external electromagnetic field in flat two-dimensional
Minkowski space-time and reduces to that in the absence of curvature.  In the
flat
limit, that interaction preserves the Poincar\'e symmetry of a non-interacting
point particle and similarly the {\it covariantly\/}
constant field $\sqrt{-g}\,\epsilon_{\mu\nu}{\cal B}$ respects general
covariance.  Both terms arise naturally in a gauge theoretical formulation;
this will be shown in Section~IV.

The additional forces can be derived from additional contributions to the
matter gravity action, containing suitable vector potentials contracted with
$\dot x^\mu$, as is clear from the electromagnetic analogy.  The action for
the force proportional to curvature evidently involves the spin connection, see
(1.8).  Construction of the action for the covariantly constant force is
geometrically subtle.$^2$ Consider the volume two-form, ${\it vol}
\equiv -{1\over 2}
\sqrt{-g}\,\epsilon_{\mu\nu}\,dx^\mu\,dx^\nu$, which may also be expressed in
terms of the {\it Zweibein\/} as $-{1\over 2} \epsilon_{ab} e^a_\mu e^b_\nu
\, dx^\mu\, dx^\nu$.  Since ${\it vol}$ is closed, $d({\it vol})=0$, it is
locally exact.
$${\it vol} = da\eqno(2.\hbox{A}7\hbox{a})$$
Equation (2.A7a) defines a one-form whose components are also seen to satisfy
$$\partial_\mu a_\nu-\partial_\nu a_\mu + \sqrt{-g}\,\epsilon_{\mu\nu}=0
\eqno(2.\hbox{A}7\hbox{b})$$
Since the right-hand side of (2.A7b) is a tensor, $a_\mu$ can be taken as a
vector. The covariant action, whose variation leads to the additional force in
(2.A6),
is now constructed from $\omega_\mu$ and $a_\mu$.
$$I_{\cal F} = - \int d\tau\,\dot x^\mu(\tau) \biggl(
{\cal A}\ \omega_\mu \bigl( x(\tau)\bigr)+
{\cal B}\ a_\mu \bigl( x(\tau)\bigr) \biggr)  \eqno(2.\hbox{A}8)$$
Under coordinate redefinition, $\omega_\mu$, $a_\mu$ and $x^\mu(\tau)$ change,
and it is straightforwardly verified that $I_{\cal F}$ is a scalar.  Under a
tangent space
Lorentz transformation
(1.12), $\omega_\mu \to \omega_\mu+\partial_\mu\alpha$, see
(1.14), and $a_\mu$ is unchanged; however, the defining equations
(2.\hbox{A}7) leave
a gauge ambiguity.
$$a_\mu\longrightarrow a_\mu +\partial_\mu \beta\eqno(2.\hbox{A}9)$$
But it is seen
that (2.\hbox{A}8) changes only by end-point contributions under the gauge
transformations (1.14), (2.A9), and the equation of motion (2.A6)
is gauge invariant.  Also it
is local, even though local expressions for $a_\mu$ and $I_{\cal F}$ are not
available.

Next we compute the covariantly conserved energy-momentum tensor
$T^{\alpha\beta}$, which is a functional of the matter variable $x(\tau)$ and
a function of the field argument $x$.
$$T^{\alpha\beta} \Bigl( x(\tau)|x\Bigr) = - {2\over \sqrt{- g(x)}} \
{\delta\over \delta g_{\alpha\beta} (x)} \left( I_m + I_{\cal F}
\right)\eqno(2.\hbox{A}10)$$
The variation of $I_m$ produces the conventional free-particle energy-momentum
tensor.
$$-{2\over \sqrt{-g(x)}} \  {\delta\over\delta g_{\alpha\beta}(x)} I_m =
{1\over \sqrt{-g(x)}}\int {d\tau\over N(\tau)} \dot x^\alpha(\tau) \dot
x^\beta(\tau)\  \delta^2\! \Bigl( x - x(\tau)\Bigr) \eqno(2.\hbox{A}11)$$
When varying $\omega_\mu$ in $I_{\cal F}$ with respect to $g_{\alpha\beta}$,
we use (1.4) for the {\it Zweibein\/}-dependence of $\omega_\mu$, and (1.1)
for the relation between the {\it Zweibein\/} and the metric tensor.  But
evaluating the metric variation of $a_\mu$, which is also present in  $I_{\cal
F}$, is problematic in the absence of an explicit formula for $a_\mu$.
However, we can fix this contribution to the energy-momentum tensor by
requiring its covariant conservation.  In this way $T^{\alpha\beta}$ is found
to be
$$\eqalign{
T^{\alpha\beta}\Bigl( x(\tau)|x\Bigr) &= {1\over \sqrt{-g(x)}}\int {d\tau\over
N(\tau)} \dot x^\alpha(\tau) \dot x^\beta(\tau) \,\delta^2\Bigl( x -
x(\tau)\Bigr) \cr
&+ {{\cal A}\over 2\sqrt{-g(x)}} \Biggl(\epsilon^{\alpha\gamma} D_\gamma
j^\beta\Bigl(x(\tau)|x\Bigr) + \epsilon^{\beta\gamma} D_\gamma j^\alpha \Bigl(
x(\tau)|x\Bigr)\Biggr) - {1\over 2}\lambda \Bigl( x(\tau)|x\Bigr)
g^{\alpha\beta} (x)
\cr}\eqno(2.\hbox{A}12)$$
Here $D_\gamma$ is the space-time covariant derivative, while $j^\mu$ is the
current,
$$j^\mu\Bigl( x(\tau)|x\Bigr) = {1\over \sqrt{-g(x)}} \int d\tau\, \dot
x^\mu(\tau) \,\delta^2\Bigl( x - x(\tau)\Bigr) \eqno(2.\hbox{A}13)$$
which is covariantly conserved.
$$D_\mu j^\mu = {1\over \sqrt{-g}}\, \partial_\mu \sqrt{-g}\,j^\mu = 0
\eqno(2.\hbox{A}14)$$
The last contribution to (2.A12) comes from varying $- {\cal B} \int d\tau\,
\dot x^\mu a_\mu$.  In view of (2.A6), $T^{\alpha\beta}$
will be covariantly conserved
$$D_\alpha T^{\alpha\beta} = 0 \eqno(2.\hbox{A}15)$$
if $\lambda$ satisfies
$${1\over 2}\ {\partial\lambda\over\partial x^\mu} \Bigl( x(\tau)|x\Bigr)= -
{\cal B} \sqrt{ - g(x)}\, \epsilon_{\mu\nu}j^\nu
\Bigl(x(\tau)|x\Bigr) \eqno(2.\hbox{A}16)$$
Equation (2.A16) is solved by
$$\lambda \Bigl( x(\tau)|x\Bigr) = -2{\cal B} \int^x dy^{\mu}
 \sqrt{ - g(y)}\, \epsilon_{\mu\nu}j^\nu \Bigl( x(\tau)|y\Bigr)
\eqno(2.\hbox{A}17)$$
Because $j^\mu$ is covariantly conserved
the expression (2.A17) for $\lambda$ depends {\it locally\/} on $x$, {\it
i.e.\/} it is path-independent.  This is established by evaluating the line
integral over the closed contour that describes the possible difference
between two different evaluations of $\lambda$ along two different paths.
Alternatively in view of (2.A14),  we may write,
$$j^\mu ={1\over\sqrt{\pi}}
{\epsilon^{\mu\nu}\over \sqrt{-g}} \partial_\nu \varphi\eqno(2.\hbox{A}18)$$
and then (2.A17)
shows that $\lambda = -{2\over\sqrt{\pi}}{\cal B}\varphi$.
When the current has the explicit
form (2.A13), the expressions (2.A16), (2.A17) or (2.A18)
may be easily evaluated in the parametrization
 $x^0 (\tau) = \tau$, and one finds
$$\lambda = - {\cal B}\varepsilon \Bigl( x^1 - x^1(t)\Bigr) +
\lambda_0=-{2\over\sqrt{\pi}}{\cal B} \varphi
\eqno(2.\hbox{A}19)$$
where $t\equiv x^0$ and
$\lambda_0$ is constant.  This is a cosmological ``constant'' that jumps
by $2{\cal B}$ as the particle's trajectory is crossed, a property that is
independent of the above parametrization choice.

The final formula for the covariantly conserved energy-momentum tensor that
follows from (2.A12), (2.A18) and (2.A19) is
$$\eqalign{T^{\alpha\beta}\big(x(\tau)|x\bigr) &= {1\over \sqrt{-g(x)}}\int
{d\tau\over N(\tau)} \dot x^\alpha(\tau) \dot x^\beta(\tau)\;\delta^2\!\bigl( x
-
x(\tau)\bigr) \cr
&+{{\cal A}\over \sqrt{\pi}} \left( D^\alpha D^\beta - g^{\alpha\beta} D_\mu
D^\mu\right) \varphi \bigl( x(\tau)|x\bigr) + {{\cal B}\over \sqrt{\pi}}
\varphi \bigl( x(\tau)|x\bigr) g^{\alpha\beta} (x)
\cr}\eqno(2.\hbox{A}20)$$
This is local in $x$, just like the equation of motion.

The trace of the energy momentum tensor reads
$$\eqalign{
T^\alpha{}_\alpha \Bigl(x(\tau)|x\Bigr) &= {m^2\over \sqrt{-g(x)}} \int
d\tau\,N(\tau)\; \delta^2\!\Bigl( x - x(\tau)\Bigr) \cr
&- {{\cal A} \over\sqrt{\pi}} D_\alpha
D^\alpha\varphi \Bigl( x(\tau)|x\Bigr) + {2{\cal B}\over\sqrt{\pi}}\varphi
\Bigl( x(\tau)|x\Bigr) \cr}\eqno(2.\hbox{A}21)$$
In conformally flat
coordinates $g_{\alpha\beta} = e^\sigma h_{\alpha\beta}$ and in
the parametrization $x^0(\tau) = \tau$,
$$\eqalign{ {1\over \sqrt{\pi}} D_\alpha D^\alpha \varphi
&= {1\over 2} e^{-\sigma}\square\,\varepsilon\bigl(x^1-x^1(t)\bigr) \cr
&= - e^{-\sigma} \left\{ a\  \delta \Bigl( x^1 -
x^1(t)\Bigr) + (1-v^2)\;\delta'\!\Bigl(x^1 - x^1(t)\Bigr) \right\}\cr}
\eqno(2.\hbox{A}22)$$
where $v = \dot x^1(t)$, $a=\ddot x^1(t)$.  For massless particles on a line,
 $v^2=1$, $a=0$ and ${1\over\sqrt{\pi}} D_\alpha D^\alpha\varphi$ vanishes,
leaving in the trace only the cosmological term.
$$T^\alpha{}_\alpha \Bigl( x(\tau)|x\Bigr)\bigg|_{m=0} =
-\lambda = {2\over\sqrt{\pi}}{\cal B}\varphi  = {\cal B}\varepsilon
\bigl(x^1-x^1(t)\bigr)-\lambda_0 \eqno(2.\hbox{A}23)$$
Note that the axial vector current $j^\mu_5$, dual to the vector current,
$$j^\mu_5 = {1\over \sqrt{-g}}\, \epsilon^{\mu\nu} j_\nu = g^{\mu\nu} {1\over
\sqrt{\pi}} \partial_\nu \varphi \eqno(2.\hbox{A}24\hbox{a})$$
occurs naturally in several of the above formulas, for example (2.A17).  In
the massless case, it also is conserved,
$$D_\mu j^\mu_5 = {1\over\sqrt{\pi}} D_\mu D^\mu\varphi = 0
\eqno(2.\hbox{A}24\hbox{b})$$
as is seen from (2.A22).

For later use, we observe that the total particle
action may also be presented in first-order form.
$$I_{\rm particle}
= \int d\tau \left( p_a e^a_\mu \dot x^\mu + {1\over 2}N\left( p^2 -
m^2\right) \right) + I_{\cal F}  \eqno(2.\hbox{A}25)$$
Upon varying and eliminating $p_a= - e^a_\mu \dot x^\mu/N$ and
$N= \sqrt{\dot x^\mu
g_{\mu\nu} \dot x^\nu}/m$, (2.A1) and (2.A8) are regained.

In the absence of gravity, $e^a_\mu = \delta^a_\mu$,
$\omega_\mu=0$, $a_\mu = {1\over 2} \epsilon_{\mu\nu} x^\nu$ and (2.A25)
reduces to
$$I^{\rm flat}_{\rm particle} = \int d\tau \left( \left( p_a -
{1\over 2}{\cal B}\epsilon_{ab}
x^b\right)\dot x^a + {1\over 2}N\left( p^2 - m^2\right)\right)
\eqno(2.\hbox{A}26)$$
so that $p_a - {{\cal B}\over 2}\epsilon_{ab} x^b$ and $x^a$ are
canonically conjugate.  Moreover, under space-time
translations $\delta x^a = x^a_0$,
$\delta p_a =0$, $\delta N=0$, $I^{\rm flat}_{\rm particle}$
changes by end-point contributions and
the conserved energy and momentum
$$P_a = p_a - {\cal B} \epsilon_{ab} x^b\eqno(2.\hbox{A}27)$$
possess non-vanishing bracket.
$$\left[ P_a, P_b\right] = {\cal B}\epsilon_{ab} \eqno(2.\hbox{A}28)$$
Infinitesimal space-time rotations $\delta x^a = \alpha
\epsilon^a{}_b x^b$, $\delta
p_a =\alpha \epsilon_a{}^b p_b$, $\delta N=0$ leave
$I^{\rm flat}_{\rm particle}$
invariant and lead to the
Lorentz generator
$$J = x^a \epsilon_a{}^bp_b- {1\over 2}{\cal B} x^2 = x^a\epsilon_a{}^b P_b
+ {1\over 2}{\cal B} x^2 \eqno(2.\hbox{A}29)$$
with bracket
$$\left[ P_a,J\right] = \epsilon_a{}^b P_b \eqno(2.\hbox{A}30)$$

In flat space-time, the energy-momentum tensor (2.A20) becomes
$$\eqalign{T^{\alpha\beta}_{\rm flat} \bigl( x(\tau)|x\bigr)
&= m \int d\tau {\dot x^\alpha(\tau) \dot x^\beta(\tau) \over \sqrt{\dot
x^2(\tau)}} \ \delta^2\!\bigl( x - x(\tau)\bigr) \cr
&+ {{\cal A} \over \sqrt{\pi}}
\left( \partial^\alpha \partial^\beta - h^{\alpha\beta} \square\right)
\varphi\bigl(x(\tau)|x\bigr)
+ {{\cal B}\over \sqrt{\pi}} \varphi\bigl( x(\tau)|x\bigr)
h^{\alpha\beta} \cr}\eqno(2.\hbox{A}31)$$
The next-to-last term corresponds to the curvature-dependent
force; even in the absence of curvature it contributes to the energy-momentum
tensor an ``improvement'' term familiar from conformally invariant coupling.
Observe further that $\int^\infty_{-\infty}dx^1\,T^{00}_{\rm flat}$
coincides with the
energy obtained from (2.A27), apart from an infinite constant proportional to
$\lambda$.  However, the spatial integral of $T^{01}_{\rm flat}$ is not the
momentum;
$\int^\infty_{-\infty}dx^1\,T^{01}_{\rm flat}$ is not time-independent even
though
$T^{\alpha\beta}_{\rm flat}$ is conserved.
This is because $T^{11}_{\rm flat}$ remains non-vanishing
at $x^1=\pm\infty$ owing to the last term in (2.A31). To achieve a
time-independent quantity and to obtain the correct momentum from the
energy-momentum tensor, one must add $t\int^\infty_{-\infty}dx^1{d\over dx^1}
T^{11}_{\rm flat}$ to $\int^\infty_{-\infty}dx^1 T^{01}_{\rm flat}$.
Similarly, construction of
the Lorentz generator from the energy-momentum tensor, begins with
$t\int^\infty_{-\infty}dx^1 T^{01}_{\rm flat} - \int^\infty_{-\infty} dx^1\,x^1
T^{00}_{\rm flat}$.  As with the momentum,
this  is not time-independent, even though $T^{\alpha\beta}_{\rm flat}$ is
conserved and symmetric, owing to slow drop-off of $T^{11}_{\rm flat}$. To
remedy this
one adds ${1\over 2}t^2 \int^\infty_{-\infty}dx^1\, {d\over dx^1}T^{11}_{\rm
flat}$, and $J$ of (2.A29) is then  reproduced,
apart from the finite constant ${\cal A}$
coming from the improvement and an infinite constant coming from the
last term in (2.A31).

In the subsequent development it becomes
convenient to replace $p_a$ by $\epsilon_a{}^b p_b$, whereupon (2.A25) is
replaced by
$$I_{\rm particle}
 = \int d\tau \left( p_a \left( - \epsilon^a{}_b e^b_\mu \dot x^\mu\right)
- {1\over 2} N \left( p^2 + m^2\right) -  \left( {\cal A}\ \omega_\mu +
{\cal B}\ a_\mu\right) \dot x^\mu\right) \eqno(2.\hbox{A}32)$$
\goodbreak
\bigskip
\noindent{\bf II.B\quad Fermi Fields}
\medskip
\nobreak
The equation for a Fermi field with mass $m$ propagating in an external
gravitational field
$$\eqalign{&i\gamma^\mu\left(
\partial_\mu - {i\over 2} \omega_\mu\gamma_5\right) \psi - m\psi = 0 \cr
&\gamma^\mu\equiv  E^\mu_a\gamma^a =  {1\over\sqrt{-g}} \epsilon^{\mu\nu}
e^a_\nu\epsilon_{ab}\gamma^b \cr}\eqno(2.\hbox{B}1)$$
can be obtained by varying with respect to $\bar{\psi}$  the following
Lagrange density.
$${\cal L} =\sqrt{-g}\left( {i\over 2}\bar{\psi}
\gamma^\mu \uprightarrow{\partial}_\mu
\psi - {i\over 2}\bar{\psi} \upleftarrow{\partial}_\mu \gamma^\mu\psi -
m\bar{\psi}\psi\right) \eqno(2.\hbox{B}2)$$
[The spin connection, in the form
(1.4), enters the equation of motion (2.B1) when upon variation of $\bar{\psi}$
in ${\cal L}$ the derivative
is moved from $\bar{\psi}$ to $\sqrt{-g}\,\gamma^\mu$; see (1.4).]
Therefore the fermion action $I_{\rm Fermi}$, including interaction with
the covariantly constant and curvature dependent forces, is
$$\eqalign{I_{\rm Fermi} = \int d^2x\,\sqrt{-g} &\biggl\{
{i\over 2} \bar{\psi} \gamma^\mu\left( \uprightarrow{\partial}_\mu
+ i \left( {\cal A}\ \omega_\mu + {\cal B}\ a_\mu\right)\right) \psi \cr
&- {i\over 2} \bar{\psi}\left(
\upleftarrow{\partial}_\mu - i  \left( {\cal A}\ \omega_\mu + {\cal B}\
a_\mu\right)\right)
\gamma^\mu \psi - m \bar{\psi}\psi \biggr\}\cr} \eqno(2.\hbox{B}3)$$
and the equation of motion reads
$$i\gamma^\mu
\left( \partial_\mu - {i\over 2}\omega_\mu \gamma_5 + i \left(
{\cal A}\ \omega_\mu + {\cal B}\
a_\mu\right)\right) \psi - m\psi = 0 \eqno(2.\hbox{B}4)$$
Evidently under a gauge transformations (1.14) and (2.A9)
of $\omega_\mu$ and $a_{\mu}$, $\psi$ must now transform as
$$\psi \to e^{{i\over 2}\alpha\gamma_5 - i  \left({\cal A}\alpha+{\cal B}\beta
\right)}\psi \eqno(2.\hbox{B}5)$$
to maintain gauge invariance, compare (1.15).

The energy-momentum tensor is found by first computing the Lorentz tensor.
$$T^\alpha_a = - {1\over\sqrt{-g}}\ {\delta I_{\rm Fermi}\over \delta
e^a_\alpha}
\eqno(2.\hbox{B}6)$$
Then
$$T^{\alpha\beta} = {1\over 2}\left( T^\alpha_a E^\beta_b + T^\beta_a
E^\alpha_b\right) h^{ab} \eqno(2.\hbox{B}7)$$
Carrying out the indicated variation gives in complete analogy with (2.A12),
$$\eqalign{ T^{\alpha\beta} &= {i\over 4}  \bar{\psi}\left(
\gamma^\alpha g^{\beta\mu}+\gamma^\beta g^{\alpha\mu}\right)
\left( \uprightarrow{\partial}_\mu + i  \left( {\cal A}\ \omega_\mu +{\cal B}\
a_\mu\right)\right) \psi \cr
&- {i\over 4} \bar{\psi} \left( \upleftarrow{\partial}_\mu
- i \left( {\cal A}\ \omega_\mu +{\cal B}\
a_\mu\right)\right) \left(\gamma^\alpha g^{\beta\mu}
 +\gamma^\beta g^{\alpha\mu}\right) \psi \cr
&+{{\cal A} \over 2} \left( {1\over \sqrt{-g}}
\epsilon^{\alpha\gamma}D_\gamma j^\beta +
{1\over \sqrt{-g}}\epsilon^{\beta\gamma} D_\gamma
j^\alpha\right)- {1\over 2}\lambda
g^{\alpha\beta} \cr}\eqno(2.\hbox{B}8)$$
where the covariantly conserved current is now
$$j^\mu = \bar{\psi}\gamma^\mu \psi \eqno(2.\hbox{B}9)$$
and $\lambda$ must again satisfy
$${1\over 2}\ {\partial\lambda\over \partial x^\mu} = - {\cal
B} \sqrt{-g}\,\epsilon_{\mu\nu}j^\nu  \eqno(2.\hbox{B}10)$$
so that $T^{\alpha\beta}$ be conserved.  The solution
$$\lambda = -2{\cal B} \int^x dy^{\mu} \sqrt{-g}\,\epsilon_{\mu\nu}j^\nu
\eqno(2.\hbox{B}11)$$
remains path-independent thanks to the covariant conservation of $j^\mu$, but
the integral cannot be further evaluated.  However, if we use (2.A18), which
here is viewed as a bosonization formula,
$\lambda$ is again proportional to $\varphi$
and the energy-momentum tensor is as in
(2.A20), except the kinetic term is now constructed from the Fermi fields.

The trace of the energy momentum tensor reads
$$T^\alpha{}_\alpha = m\bar{\psi}\psi - {\cal A} \;  D_\mu j^\mu_5 -
\lambda\eqno(2.\hbox{B}12)$$
where the axial current, dual to the vector current (2.B9),
$$j^\mu_5 = \bar{\psi} i \gamma_5 \gamma^\mu \psi \eqno(2.\hbox{B}13)$$
possesses the covariant divergence $2m\bar{\psi}\gamma_5 \psi$, so that
$$T^\alpha{}_\alpha = m \bar{\psi} \left( 1 - 2{\cal A} \gamma_5 \right)
\psi - \lambda \eqno(2.\hbox{B}14)$$
In the massless case one is simply left with $T^\alpha{}_\alpha=-\lambda$.  Of
course the above does not include the trace and chiral anomalies which are
quantum effects, see below.

In the massless case we can evaluate the fermion determinant by making use of
known results,$^8$
which are given in Riemannian space.
Adjustments in our formulas necessitated by a positive signature metric are:
$h_{ab}$ is now $\diag(1,1)$, while the {\it Zweibein\/}, spin-connection
and curvature are related by
$$\eqalignno{\omega_\mu &= - {1\over \sqrt{g}}
\epsilon^{\alpha\beta} e^a_\mu h_{ab}
\partial_\alpha e^b_\beta &(1.14') \cr
\partial_\mu \omega_\nu - \partial_\nu \omega_\mu &= {1\over
2}\sqrt{g}\,\epsilon_{\mu\nu} R &(1.8')\cr}$$
The $\gamma$ matrices fulfill the Euclidean Clifford algebra.

According to Ref.~[8], the determinant of the massless, two-dimensional Dirac
operator in curved space and in the presence of an external Abelian gauge
field $A_\mu$ is
$$\eqalign{W(E,A)
&\equiv \ln \det \left[i\gamma^a E^\mu_a \left( \partial_\mu -
{i\over 2}\omega_\mu \gamma_5 + i A_\mu\right)\right] \cr
&=-{1\over 12} W_g (\omega) + W_g (A) \cr}\eqno(2.\hbox{B}15)$$
The functional $W_g(v)$ of the vector field $v_\mu$ is defined in terms of the
inverse Laplacian $\nabla^{-2}_g$,
$$\eqalign{g^{\mu\nu} (x) D^x_\mu D^x_\nu \nabla^{-2}_g (x,y)
&= {1\over \sqrt{g}}\delta^2(x-y)\cr
\nabla^{-2}_g (x,y) &= \nabla^{-2}_g (y,x)\cr} \eqno(2.\hbox{B}16)$$
and is given by
$$W_g (v) = {1\over 2\pi} \int d^2x\,d^2y\,\epsilon^{\mu\nu} \partial_\mu
v_\nu (x) \nabla^{-2}_g(x,y) \epsilon^{\alpha\beta} \partial_\alpha
v_\beta(y) \eqno(2.\hbox{B}17)$$
Local terms are adjusted in (2.B15) to ensure invariance against coordinate
transformations, and also against Abelian gauge transformations of
$\omega_\mu$ and $A_\mu$.

For us, $\epsilon^{\mu\nu} \partial_\mu \omega_\nu =  {1\over 2}\sqrt{g}\,R$,
and since $A_\mu = {\cal A}\omega_\mu + {\cal B}a_\mu$, $\epsilon^{\mu\nu}
\partial_\mu A_\nu =  \sqrt{g}\, \left( {{\cal A}\over 2} R+{\cal
B}\right)$.  Note that ${\cal B}^2\int d^2x \sqrt{g(x)}\, d^2y \sqrt{g(y)}\,
\nabla^{-2}_g(x,y)$ may diverge, {\it e.g.\/} in flat space.

The vacuum expectation value of the energy-momentum tensor is obtained from
$W(E,A)$ by varying with respect to gravitational variables.  One finds
$$\eqalign{\ll T_{\alpha\beta}\rr &= - {1\over 48\pi} \left( \partial_\alpha
\sigma\partial_\beta\sigma - {1\over 2} g_{\alpha\beta} \partial_\mu\sigma
\partial^\mu\sigma\right)
+ {1\over 24\pi} \left( D_\alpha D_\beta - g_{\alpha\beta} D_\mu D^\mu
\right)\sigma \cr
&- \left( \partial_\alpha \varphi \partial_\beta \varphi - {1\over 2}
g_{\alpha\beta} \partial_\mu \varphi \partial^\mu\varphi\right)
+ {{\cal A}\over \sqrt{\pi}} \left( D_\alpha D_\beta
- g_{\alpha\beta} D_\mu D^\mu\right) \varphi \cr
&- {{\cal A}\over 2\sqrt{\pi}} \left(\partial_\alpha \sigma \partial_\beta
\varphi + \partial_\beta\sigma\partial_\alpha \varphi - g_{\alpha\beta}
\partial_\mu \sigma
\partial^\mu\varphi\right)  + {{\cal B}\over \sqrt{\pi}} \varphi
g_{\alpha\beta}\cr}  \eqno(2.\hbox{B}18)$$
In order to present $\ll T_{\alpha\beta}\rr$ in
local form we have introduced the fields
$$\sigma\equiv - \nabla^{-2}_g R \eqno(2.\hbox{B}19)$$
--- a field that
coincides with the conformal factor in conformal coordinates --- and
$$\varphi\equiv {1\over \sqrt{\pi}} \nabla^{-2}_g {1\over\sqrt{g}}
\epsilon^{\mu\nu} \partial_\mu A_\nu\eqno(2.\hbox{B}20)$$
This energy-momentum tensor is covariantly conserved but exhibits a trace
anomaly.
$$\ll T^\alpha{}_\alpha\rr = {1\over 24\pi} \left( 1 - 12{\cal A}^2\right) R -
{1\over\pi} {\cal A}{\cal B} + {2\over\sqrt{\pi}}
{\cal B} \varphi \eqno(2.\hbox{B}21)$$
The expressions (2.B18) and (2.B21) are similar to (2.A20) and (2.A21).  They
can be written as (2.A12) or (2.B8), (2.B12) if we introduce the conserved
current,
$$\eqalign{\ll j^\mu\rr &= - {1\over\sqrt{g}}\  {\delta W_g\over \delta A_\mu}
\cr
&= - {1\over \sqrt{\pi}}\  {\epsilon^{\mu\nu}\over\sqrt{g}} \partial_\nu
\varphi\cr}\eqno(2.\hbox{B}22)$$
and its dual, the axial current.
$$\ll j^\mu_5\rr = {1\over \sqrt{g}} \epsilon^{\mu\nu} j_\nu = {1\over
\sqrt{\pi}} g^{\mu\nu} \partial_\nu \varphi\eqno(2.\hbox{B}23)$$
The latter has an anomalous divergence
$$D_\mu \ll j^\mu_5\rr = {1\over\pi} \left( {{\cal A}\over 2} R + {\cal
B}\right)\eqno(2.\hbox{B}24)$$
which appears explicitly in the trace equation (2.B21);
$$\eqalignno{\ll T^\alpha{}_\alpha\rr &= {1\over 24\pi} R - {\cal A} D_\mu \ll
j^\mu_5\rr - \lambda &(2.\hbox{B}25\hbox{a}) \cr
{1\over 2}\partial_\mu \lambda &= - {\cal B} \sqrt{g}\,\epsilon_{\mu\nu} \ll
j^\nu\rr &(2.\hbox{B}25\hbox{b}) \cr}$$
compare to (2.B12), (2.B10).

We now return to Minkowski signature.
\goodbreak
\bigskip
\noindent{\bf III.\quad EXTENDED POINCAR\'E GROUP}
\medskip
\nobreak
The somewhat haphazard introduction of the additional interactions becomes
rationalized in a group theoretical description.  We now discuss the extended
Poincar\'e group that we employ.

The Poincar\'e group consists of two translations (parameters $\theta^a$) and
single (Lorentz) rotation $\Lambda$ [parameter $\alpha$, see (1.12)] with
composition law
$$\left( \theta_1,\Lambda_1\right) \circ \left( \theta_2,\Lambda_2\right) =
\left( \theta_1 + \Lambda_1  \theta_2, \Lambda_1\Lambda_2\right) \eqno(3.1)$$
Thus $(\theta,\Lambda)$ corresponds to $e^{\theta^a{\cal P}_a}\,e^{\alpha
{\cal J}}$, where
generators ${\cal P}_a$ effect translations  and ${\cal J}
$ the infinitesimal rotation.  We
postulate that the generator
 Lie algebra follows (2.A28), (2.A30), {\it i.e.\/}, there is
an extension.  Henceforth we scale ${\cal B}$ to unity.
$$\eqalignno{\left[{\cal P}_a, {\cal P}_b\right] &= \epsilon_{ab} {\cal I}
\ \ ,&(3.2) \cr
\left[{\cal P}_a, {\cal J}\right] &= \epsilon_a{}^b {\cal P}_b &(3.3)\cr}$$
${\cal I}$ is a central element, commuting with ${\cal P}_a$ and ${\cal J}$.
Note that ${\cal J}$ may be
supplemented by an arbitrary multiple of ${\cal I}$. The algebra is solvable.
[The algebra is similar to that of the
harmonic oscillator group $Os(1)$, but in our realization we do not impose
reality/Hermiticity requirements on the generators.$^9$]

The effect of the center is to modify the composition law for group elements,
represented by exponentiated generators.
$$\left( \theta_1, \Lambda_1\right) \circ \left(\theta_2, \Lambda_2\right) =
e^{{1\over 2}\theta^a_1\epsilon_{ab} \left(
\Lambda_1\theta_2\right)^b {\cal I}}\left( \theta_1 + \Lambda_1 \theta_2,
\Lambda_1
\Lambda_2\right)\eqno(3.4)$$
In order to obtain a faithful representation, we extend the Poincar\'e group
with a $U(1)$ factor, generated by ${\cal I}$ and  parameter $\beta$.  Upon
defining
group elements
$${\cal U} \left( \theta, \Lambda,\beta\right) = e^{\theta^a
{\cal P}_a} e^{\alpha {\cal J}} e^{\beta {\cal I}} \eqno(3.5)$$
we verify that
$${\cal U}\left( \theta_1, \Lambda_1, \beta_1\right) {\cal U} \left(
\theta_2, \Lambda_2, \beta_2\right) = {\cal U}\left( \theta_3, \Lambda_3,
\beta_3\right) \eqno(3.6\hbox{a})$$
with
$$\eqalign{
\theta_3 &= \theta_1 + \Lambda_1\theta_2\cr
\Lambda_3 &= \Lambda_1 \Lambda_2\cr
\beta_3 &= \beta_1 + \beta_2 + {1\over 2} \theta^a_1\epsilon_{ab}
\left( \Lambda_1
\theta_2\right)^b \cr}\eqno(3.6\hbox{b})$$
In this way one is led to a four-parameter group --- the extended
$(1+1)$-dimensional Poincar\'e group.

In a covariant notation, we call the generators ${\cal Q}_A$,
where $A$ takes the four
values ($a,2,3)$, ${\cal Q}_A=({\cal P}_a,{\cal J},{\cal I})$.
The algebra is four-dimensional.
$$\left[{\cal Q}_A,{\cal Q}_B\right] = f_{AB}{}^C {\cal Q}_C \eqno(3.7)$$
Also we verify
$${\cal U}^{-1}{\cal Q}_A {\cal U} = {\cal Q}_B
\left( U^{-1}\right)^B{}_A \eqno(3.8)$$
with $U$ the $4\times 4$ matrix
$$U^A{}_B = \left( \matrix{ \Lambda^a{}_b & - \epsilon^a{}_c \theta^c & 0
\cr\noalign{\vskip 0.2cm}
0 & 1 & 0 \cr\noalign{\vskip 0.2cm}
\theta^c \epsilon_{cd} \Lambda^d{}_b &  -{1\over 2} \theta^c \theta_c
& 1 \cr}\right)\eqno(3.9)$$
\goodbreak
\bigskip
\noindent{\bf A.\quad Adjoint Representation}
\medskip
\nobreak
$U$ gives the adjoint representation, where the generators are
represented by $Q_C$,
$$\left( Q_C\right)^A{}_B = f_{CB}{}^A \eqno(3.10\hbox{a})$$
explicitly
$$\left( P_c\right)^A{}_B = \left( \matrix{ 0 & -\epsilon^a{}_c &
0 \cr\noalign{\vskip 0.2cm}
0 & 0 & 0 \cr\noalign{\vskip 0.2cm}
\epsilon_{cb} & 0 & 0 \cr}\right) \ \ ,\quad
\left( J\right)^A{}_B = \left( \matrix{ \epsilon^a{}_b & 0 & 0
\cr\noalign{\vskip 0.2cm}
0 & 0 & 0 \cr\noalign{\vskip 0.2cm}
0 & 0 & 0 \cr}\right) \ \ ,\quad
\left( I\right)^A{}_B = 0 \eqno(3.10\hbox{b})$$
It is also true that $U^A{}_B$ coincides with ${\cal U}$ of
(3.5), with the exponential generators taken in the adjoint representation
(3.10).  [Our convention is that non-script letters denote group and algebra
elements in a specific representation.]

In the usual way, we can define contravariant four-vectors
$\xi^A$ that transform
according to the adjoint transformation,
$$\xi^A\longrightarrow \left( U^{-1}\right)^A{}_B \xi^B\eqno(3.11\hbox{a})$$
or infinitesimally
$$\delta \xi^A = f_{BC}{}^A \xi^B\Theta^C \eqno(3.11\hbox{b})$$
where $\Theta^A= \left(\theta^a,\alpha,\beta\right)$.
Similarly there are covariant four-vectors with coadjoint transformation.
$$\eqalignno{\eta_A&\longrightarrow \eta_B U^B{}_A
&(3.12\hbox{a})\cr
\delta \eta_A &= - f_{AB}{}^C \Theta^B \eta_C &(3.12\hbox{b}) \cr}$$
Evidently $\eta_A\xi^A$ is a scalar invariant.

An invariant mixed tensor is provided by the structure constants.
$$f_{AB}{}^C \longrightarrow \left( U^{-1}\right)^C{}_{C'}
f_{A'B'}{}^{C'}
U^{A'}{}_A U^{B'}{}_B = f_{AB}{}^C \eqno(3.13)$$
Since the algebra is solvable, there also exist invariant four-vectors: upon
defining $i^A = \left(\matrix{\phantom{-}0\cr\phantom{-}0\cr-1\cr}\right)$ and
$i_A=\left(0,1,0\right)$, $i_Ai^A=0$, we see that
$$\eqalignno{\left( U^{-1}\right)^A{}_B\; i^B &= i^A &(3.14\hbox{a}) \cr
i_B U^B{}_A &= i_A &(3.14\hbox{b})\cr}$$
One understands the existence of such invariant vectors because equivalent to
(3.14) are the statements
$$\eqalignno{i^A f_{AB}{}^C &= 0 &(3.15\hbox{a}) \cr
f_{AB}{}^C i_C &= 0 &(3.15\hbox{b}) \cr}$$
The first reflects the commutativity of ${\cal I}$
with all other generators, the
second is true because ${\cal J}$ is never attained by commuting two
generators.
These invariant vectors also determine the invariant Cartan--Killing metric,
$$\ll {\cal Q}_A, {\cal Q}_{A'}\rr_{C-K} =
f_{AB}{}^C f_{A'C}{}^B = 2i_A i_{A'}\eqno(3.16)$$
which is however singular, since the group is not semi-simple.

An invariant, non-singular bilinear form can be defined in the adjoint
representation. It is given by the tensor
$$h_{AB} = \left( \matrix{ h_{ab} & \phantom{-}0 & \phantom{-}0
\cr\noalign{\vskip 0.2cm}
0 & \phantom{-}0 & -1 \cr\noalign{\vskip 0.2cm}
0 & -1 & \phantom{-}0\cr}\right) \equiv \ll{\cal Q}_A,{\cal
Q}_B\rr\eqno(3.17)$$
$h_{AB}$ is used to raise and lower indices, interchanging covariant and
contravariant tensors, as for example
$$i_A = h_{AB} i^B \eqno(3.18)$$
This allows defining an invariant length,
$$\xi^A \xi_A = \xi^a \xi_a - 2\xi_2 \xi_3 \eqno(3.19)$$
and a bilinear Casimir invariant can be constructed from the generators.
$${\cal C} = {\cal Q}^A{\cal Q}_A = {\cal P}^a {\cal P}_a -2{\cal I}{\cal J}
\eqno(3.20)$$
In the adjoint representation, the Casimir is determined by the
Cartan--Killing metric.
$$C^A{}_B = 2i^A i_B\eqno(3.21)$$
When the upper index on the structure constant is lowered,
$$f_{ABC} = f_{AB}{}^{C'} h_{C'C}\eqno(3.22\hbox{a})$$
the resulting covariant tensor is antisymmetric in all three indices and
vanishes when any index equals three.  The only non-vanishing components are
permutations of
$$f_{ab2} = -\epsilon_{ab} \eqno(3.22\hbox{b})$$
and it is true that
$$f_{ABC}\,f_{A'B'C'} = -{1\over 2}
 i_A i_{A'} h_{BB'} h_{CC'} + \hbox{permutations}
\eqno(3.22\hbox{c})$$

According to (3.19), the length of a vector $\xi^A$ is invariant.
Moreover, the presence of the invariant vector $i_A$ shows that the third
component of $\xi^A$ (alternatively the last component of $\xi_A$) also is
invariant, because it can be expressed as $\xi^Ai_A=i^A\xi_A$.  In the next
Section, where we present a Poincar\'e gauge invariant and manifestly
covariant formulation of our dynamics,
we shall introduce a ``Poincar\'e coordinate''
four-vector $q^A$, with length $-2{\cal A}$ and third component set to 1.
$$q^A =\left( \matrix{ q^a\cr\noalign{\vskip 0.2cm} 1\cr\noalign{\vskip 0.2cm}
{1\over 2} q^b q_b + {\cal A}\cr}\right) \eqno(3.23)$$
Also we shall use an on-mass-shell momentum four-vector $p_A$, with vanishing
last component.
$$p_A = \left( p_a, p_2, 0\right)\ \ ,\qquad p_ap^a = - m^2 \eqno(3.24)$$
In view of the above remarks, these restrictions are invariant: $q^Ai_A = 1$,
$q^A q_A+2{\cal A}=0$; $i^A p_A =0$, $p^A p_A+m^2=0$.  It follows from (3.11)
and (3.12)
that the transformation law for $q^a$ and $p_a$ is
$$\eqalignno{q^a &\longrightarrow \left( \Lambda^{-1}\right)^a{}_b \left( q^b
+ \epsilon^b{}_c \theta^c\right) & (3.25) \cr
p_a & \longrightarrow p_b \Lambda^b{}_a  &(3.26)\cr}$$

Note that the specific translation
$${\cal T} (q)= e^{-q^a \epsilon_a{}^b {\cal P}_b}\eqno(3.27)$$
taken in the adjoint representation
$$T(q)^A{}_B = \left( \matrix{
\delta^a{}_b & -q^a & 0 \cr\noalign{\vskip 0.2cm}
0 & 1 & 0 \cr\noalign{\vskip 0.2cm}
-q_b & {1\over 2} q^c q_c & 1 \cr}\right) \eqno(3.28\hbox{a})$$
moves $q^A$ to its ``origin'' where $q^a$ vanishes.
$$T(q)^A{}_B q^B = \left( \matrix{ 0\cr1\cr
 {\cal A}\cr}\right)\eqno(3.28\hbox{b})$$

In the four-dimensional adjoint representation, the momenta are realized
non-trivially by commuting matrices and the center is realized trivially, see
(3.10).  Indeed the matrices provide a four-dimensional representation of the
non-extended Poincar\'e group.  Of course in the extended
algebra (3.2), (3.3), the center can be represented by the identity only in an
infinite-dimensional realization.  Nevertheless, we now show that it is
possible to represent
the center by a finite-dimensional identity matrix,
in a formalism where the translations have been ``neutralized.''$^{10}$
\goodbreak
\bigskip
\noindent{\bf B.\quad  Infinite-Dimensional Representation}
\medskip
\nobreak
Consider a quantity $\Phi$ transforming according to some unspecified
representation in which all generators are realized non-trivially.
$$\Phi\longrightarrow {\cal U}^{-1}\Phi \eqno(3.29)$$
Here ${\cal U}$ is as in (3.5), but realized in the above unspecified
representation.  Consider next
$$\Psi = {\cal T}(q)\Phi\eqno(3.30)$$
where ${\cal T}$ is as in (3.27), with ${\cal P}_a$ realized in the appropriate
representation.    Observe that
$\Psi$ transforms as
$$\Psi \to {\cal T}\!\left(\Lambda^{-1} (q+\epsilon\theta)\right)
{\cal U}^{-1}\Phi = {\cal T}\!\left(\Lambda^{-1}(q+\epsilon\theta)\right)
{\cal U}^{-1}{\cal T}^{-1} (q) \Psi   \eqno(3.31\hbox{a})$$
Upon combining factors, it follows that
$$\Psi\longrightarrow e^{-\alpha {\cal J}}
e^{-\left(\beta -{1\over 2} q_a
\theta^a\right){\cal I}} \Psi \eqno(3.31\hbox{b})$$
Unlike $\Phi$,
$\Psi$ does not transform in a manifestly
covariant fashion; nevertheless it enjoys a definite place in the theory:
because ${\cal P}_a$ has disappeared from the transformation law, the
remaining generators, which commute,
may be represented by finite-dimensional matrices
acting on a finite-dimensional quantity,
with ${\cal I}$ proportional to the identity matrix.
We shall show that Fermi fields behave precisely in this fashion under
extended Poincar\'e gauge transformations.  Effectively an infinite number of
components in $\Psi$ decouples, leaving a two-component Fermi field $\psi$.

$\Bigl($The phenomenon of representing a group without using the generators
whose
commutator is a $c$-number can also be seen in the Heisenberg algebra, which
forms a nilpotent subalgebra of our extended Poincar\'e algebra.$^9$  Calling
the Heisenberg generators $X$, $P$ and $\equalsone$ [analogous to the above
$iP_+$,
$iP_-$ and $-iI$] we can represent a group element $U\left(
\theta^1,\theta^2,\beta\right)$ by $e^{i(\theta^1X+\theta^2P)}
e^{i\beta\equalsone}$ and its action on a function of $x$, $\Phi(x)$, is
$\Phi(x) \mapsto {\cal U}^{-1} \Phi (x) = e^{-i\beta} e^{{i\over
2}\theta^1\theta^2} e^{-i\theta^1x} \Phi (x-\theta^2)$.  In order to remove
the action of the generators $X$ and $P$, introduce two more
variables $q^1$ and $q^2$, which also transform according to $q^1\to q^1 -
\theta^1$, $q^2 \to q^2 + \theta^2$, and define a new set of functions of $x$,
$\Psi(x)$, by $\Psi(x) \equiv e^{-i\left( q^1X-q^2P\right)} \Phi(x) =
e^{-iq^1x} e^{-{i\over 2}q^1q^2}\Phi(x+q^2)$.  It then follows that $\Psi(x)$
responds to the transformation ${\cal U}\left(\theta^1,\theta^2,\beta\right)$
without the appearance of $P$ and $X$: $\Psi(x)\mapsto{\cal U}^{-1} \Psi(x) =
e^{-i\left( \beta - {1\over 2} \left( q^1\theta^2 + q^2
\theta^1\right)\right)} \Psi(x)$.$\Bigr)$

In the infinite-dimensional representation, which we shall use below when
discussing Fermi fields before passing to the finite-dimensional $\psi$,
the $P_a$ are matrices familiar from harmonic
oscillator theory: in light-cone components $\left[\pm={1\over\sqrt{2}}(0\pm
1)\right]$
$$\eqalignno{\ll n' \left|P_+ \right|n\rr &=\epsilon
\sqrt{\nu+n}\, \delta_{n',n-1} &(3.32\hbox{a}) \cr
\ll n' \left|P_-\right| n\rr &=\epsilon
\sqrt{\nu+n+1}\,\delta_{n',n+1} &(3.32\hbox{b}) \cr}$$
with $\nu$ arbitrary, $0\le\Re\nu<1$,
and $n,n' = 0,\pm1,\pm2,\ldots$~.  Equation (3.2) is verified and
${\cal I}$ is realized by the (infinite) identity matrix, $\equalsone$,
multiplied by
an arbitrary constant $\epsilon^2$: $I = \epsilon^2\equalsone$.  The quadratic
Casimir $C$ in (3.20) will commute with the above infinite $P_\pm$
matrices only if
it is proportional to $\equalsone$.  Therefore the Lorentz generator is found
from (3.20) to be simultaneously diagonal with $I$, with which it commutes,
$$\ll n' \left|J\right|n\rr = \left(j+n\right)\delta_{n',n}\eqno(3.33)$$
where $j$ is an arbitrary number determined by $\nu$ and the Casimir.
\goodbreak
\bigskip
\noindent{\bf C.\quad Gauge Fields}
\medskip
\nobreak
Associated with our extended Poincar\'e Lie algebra, are gauge connections
$A^A_\mu$.  We define the Lie algebra valued one-form from the vector
potentials that are present in the theory.
$$A = A^A_\mu\, dx^\mu\, {\cal Q}_A
= e^a_\mu\, dx^\mu\, {\cal P}_a + \omega_\mu \, dx^\mu\,
{\cal J} + a_\mu \, dx^\mu\, {\cal I} \eqno(3.34)$$
Here $\omega_\mu$ and $a_\mu$ are independent quantities, not satisfying any
other formulas like (1.4) and (2.A7).
The gauge curvature two-form
$$F = dA + A^2 \eqno(3.35\hbox{a})$$
has components
$$F = F^A {\cal Q}_A
= \left(de^a + \epsilon^a{}_b \omega e^b\right) {\cal P}_a + d\omega\,
{\cal J} + \left( da + {1\over 2} e^a \epsilon_{ab} e^b\right) {\cal I}
\eqno(3.35\hbox{b})$$
According to (1.5) and (1.8) the gauge curvature along the translation and
rotation generators is the torsion density and scalar curvature density,
respectively, while
the gauge curvature along the central element coincides with the left side of
(2.A7a).
Evidently $F^A$ transforms as a contravariant vector, {\it i.e.\/} as in
(3.11), or $F\to {\cal U}^{-1} F\;{\cal U}$,
 as also does $A^A$ supplemented by a gauge transformation:
$A\longrightarrow {\cal
U}^{-1}A \ {\cal U} + {\cal U}^{-1} d\,{\cal U}$.  In components
$$\eqalignno{
e^a_\mu &\longrightarrow \left( \Lambda^{-1}\right)^a_{\;b} \left( e^b{}_\mu +
\epsilon^b{}_c \theta^c\omega_\mu
+\partial_\mu\theta^b\right)&(3.36\hbox{a})\cr
\omega_\mu &\longrightarrow \omega_\mu + \partial_\mu\alpha &(3.36\hbox{b})\cr
a_\mu &\longrightarrow a_\mu - \theta^a\epsilon_{ab} e^b_\mu -{1\over 2}
\theta^a\theta_a\omega_\mu + \partial_\mu\beta +
{1\over 2}\partial_\mu \theta^a \epsilon_{ab} \theta^b &(3.36\hbox{c})\cr}$$
These comprise  the previously discussed Lorentz transformation (1.13), (1.14),
and gauge transformation on $a_\mu$ (2.A9), now further
supplemented by a translation,
which also produces a local gauge transformation.

Given a covariant quantity $\Phi$, we define
the gauge covariant derivative ${\cal D}_\mu\Phi$ by
$${\cal D}_\mu \Phi \equiv \partial_\mu \Phi + A^A_\mu {\cal Q}_A \Phi
\eqno(3.37)$$
For the adjoint representation this formula reads
$$\left({\cal D}_\mu \xi\right)^A = \partial_\mu \xi^A + f_{BC}{}^A A^B_\mu
\xi^C
\eqno(3.38\hbox{a})$$
or using the Lie algebra-valued quantity
$$\eqalignno{ \xi &= \xi^A {\cal Q}_A &(3.38\hbox{b})\cr\noalign{\hbox{we
have}}
{\cal D}_\mu \xi &=
\partial_\mu \xi + \left[ A_\mu,\xi\right] &(3.38\hbox{c})\cr}$$
Of special interest is the gauge covariant derivative of the Poincar\'e
coordinate $q^A$ in (3.23).
$$\eqalignno{\left( {\cal D}_\mu
q\right)^A &=\partial_\mu q^A + f_{BC}{}^A A^B_\mu q^C= \left( \matrix{ \left(
{\cal D}_\mu q\right)^a\cr\noalign{\vskip 0.2cm} 0\cr\noalign{\vskip 0.2cm}
q_b \left( {\cal D}_\mu q\right)^b \cr}\right) &(3.39\hbox{a})
\cr\noalign{\vskip 0.2cm}
\left( {\cal D}_\mu q\right)^a &= \partial_\mu q^a + \epsilon^a{}_b \left( q^b
\omega_\mu - e^b_\mu\right) &(3.39\hbox{b}) \cr}$$
which transforms as
$$\left( {\cal D}_\mu q\right)^a\to \left(
\Lambda^{-1}\right)^a{}_b \left( {\cal D}_\mu q\right)^b \eqno(3.40)$$
[because the third component of $\left( {\cal D}_\mu q\right)^A$ vanishes].
\goodbreak
\bigskip
\noindent{\bf D.\quad Comments}
\medskip
\nobreak
We conclude with several observations.  The metric (3.17) on the algebra that
we have introduced can be generalized: one may add an arbitrary multiple of
the Cartan--Killing metric (3.16).
$$\tilde h_{AB}=h_{AB} + c\ i_A i_B
= \left( \matrix{ h_{ab} & \phantom{-}0 & \phantom{-}0
\cr\noalign{\vskip 0.2cm}
 0 & \phantom{-}c & -1 \cr\noalign{\vskip 0.2cm}
0 & -1 & \phantom{-}0\cr}\right) \eqno(3.41)$$
Evidently
$$\eqalign{
q^A_1 \tilde h_{AB} q^B_2 &= q^A_1 h_{AB} q^B_2 + c\left(q^A_1i_A\right)
\left( i_Bq^B_2\right) \cr
{\cal Q}_A\tilde h^{AB} {\cal Q}_B &= {\cal P}_a h^{ab}{\cal P}_b - 2{\cal
J}{\cal I} - c{\cal I}^2\cr}\eqno(3.42)$$
Thus, with the modified metric, the ``length'' and inner product of vectors is
shifted by an arbitrary amount.  Equivalently we see that introducing the
parameter $c$ is tantamount to shifting ${\cal J}$ by a multiple of
${\cal I}$: ${\cal J}\to {\cal J}+(c/ 2){\cal I}$.
In our application below, we deal with vectors of arbitrary
length and/or vanishing component along $i_A$, so it does not appear that the
one-parameter freedom of (3.41) adds any more generality to our theory ---
beyond the freedom that is already contained in the fact that the
Poincar\'e  coordinate $q^A$ satisfies $q^A q_A+2{\cal A}=0$, and that
${\cal J}$ may be shifted at will.   Thus we ignore the
possibility (3.41) and remain with (3.17). [Note that it was already observed,
in the end of Section~II.A, that the Lorentz generator (2.A29) differs by a
constant from the expression obtained with the energy momentum tensor.]

The metric (3.17) may be diagonalized, and it is then seen that the signature
is \hbox{$(1,-1,1,-1)$}.
Consequently our Poincar\'e group adjoint representation
also gives a representation for a subgroup of $SO(2,2)$.
The occurrence of this group in the
present context is mysterious.  To be sure,
$SO(2,2)$ is the conformal group in $(1+1)$
dimensions, and our Poincar\'e coordinate with ${\cal A}=0$ is like Dirac's
projective variable for realizing conformal transformations linearly.
However, in general ${\cal A}\not=0$, and also we do not deal exclusively with
massless particles, so there does not seem to be any actual relevance of the
conformal group.
\goodbreak
\bigskip
\hangindent=28pt\hangafter=1
\noindent{\bf IV.\quad POINCAR\'E GAUGE INVARIANCE
OF MATTER-GRAVITY\hfill\break
DYNAMICS}
\medskip
\nobreak
In this Section we present the actions (2.A32) and (2.B3) for the point
particle and the Fermi field, respectively, in a formalism that is invariant
against gauge transformations of the extended Poincar\'e group.  The gauge
invariant expressions are written both in component form and in a manifestly
covariant tensor formalism.

Our strategy for coupling matter to gravity in a gauge invariant manner
follows (a modified version of) the approach due to Grignani and Nardelli,$^7$
which we first describe in general terms.

To achieve a Poincar\'e gauge invariant description, one begins by presenting
the dynamics in terms of the {\it Zweibeine\/} $e^a_\mu$, spin connection
$\omega_\mu$ and the additional Abelian connection $a_\mu$ associated with the
center.  In this form, the action is invariant against Lorentz gauge
transformations, but not against translations since neither $e^a_\mu$ nor
$a_\mu$ are translation covariant, compare (3.36).  To achieve
gauge invariance against translations, a Poincar\'e-coordinate $q^a$ is
introduced, which transforms as in (3.25), {\it i.e.\/} as
the first two components of a four-vector
$q^A$, whose third component is 1, $q^Ai_A=1$, and whose length, $q^A
q_A = -2{\cal A}$; compare (3.23).
The Poincar\'e coordinate provides a mapping from Minkowski space to the
tangent space, {\it i.e.\/} it is a function of $x^\mu$: $q^a(x)$; in the
point-particle application $q^a$ is evaluated on the path $x^\mu(\tau)$ and so
may be taken as a function of $\tau$: $q^a\bigl(x(\tau)\bigr) \equiv
q^a(\tau)$.

The invariant action is now obtained by replacing $-\epsilon^a{}_be^b_\mu$ by
$\left( {\cal D}_\mu q\right)^a$, and ${\cal A}\omega_\mu + a_\mu$ by
$-q_A A^A_\mu + {1\over 2} q^a
\epsilon_{ab} \partial_\mu q^b$.  This renders a Lorentz gauge invariant
 action also Poincar\'e gauge
invariant because of the following two facts.  First,
$\left({\cal D}_\mu q\right)^a$ is unaffected by translations ---
it transforms solely by a Lorentz
rotation, see (3.40).  Second, $-q_AA^A_\mu +
{1\over 2} q^a \epsilon_{ab} \partial_\mu q^b$ changes by a total derivative,
$$\eqalign{-q_A A^A_\mu &+ {1\over 2} q^a \epsilon_{ab}\partial_\mu
q^b\longrightarrow \cr
-q_AA^A_\mu &+{1\over 2} q^a \epsilon_{ab} \partial_\mu q^b + \partial_\mu
\left({\cal A}\alpha+\beta - {1\over 2} q_a \theta^a\right) \cr}\eqno(4.1)$$
and this lack of invariance can be rendered innocuous.

The invariant action now depends on an additional variable --- the Poincar\'e
coordinate $q^a$.  Dynamical equations are obtained by varying {\it all\/} the
variables --- the original ones and $q^a$.  However, the invariant
content of the dynamics is not affected by the additional variable.  This is
established by  the following consideration.

Observe from (3.25) that gauge transformations shift $q^a$ by an arbitrary
amount, so that $q^a$ may be set to zero --- indeed the $T$ transformation
(3.28) accomplishes this.  At $q^a=0$, $\left({\cal D}_\mu q\right)^a$
reduces to
$-\epsilon^a{}_b e^b_\mu$, see (3.39b), while $-q_AA^A_\mu+{1\over 2}q^a
\epsilon_{ab} \partial_\mu q^b$ becomes ${\cal A}\omega_\mu+a_\mu$.
Therefore the
equations of motion, obtained by varying dynamical variables other than
$q^a$, coincide at $q^a=0$ with the equations of motion of the Lorentz
invariant theory; at the same time $q^a=0$ represents an attainable gauge
choice in
the Poincar\'e invariant theory.  It remains to examine the equation obtained
by varying $q^a$. Here one sees that at $q^a=0$ this equation is automatically
satisfied, when all the other equations of motions hold.  The proof follows
from
gauge invariance of the action:  Upon denoting all variables other than $q^a$
by
$\chi$, with infinitesimal gauge transform $\delta\chi$, and with $\delta
q^a= -\epsilon^a{}_b
q^b\alpha + \epsilon^a{}_b\theta^b$ being the infinitesimal gauge transform of
$q^a$ [see (3.25)], gauge invariance can be stated as the following property
of the total action $I_t$
$$0 = \int {\delta I_t\over \delta\chi}\delta\chi + \int {\delta I_t\over
\delta
q^a} \delta q^a = \int {\delta I_t\over\delta\chi}\delta\chi + \int {\delta
I_t\over \delta q^a}\left( - \epsilon^a{}_b q^b \alpha + \epsilon^a{}_b
\theta^b\right)\eqno(4.2\hbox{a})$$
The first term in the last equality
vanishes when equations of motion for the $\chi$ variables are satisfied, while
the last term leaves at $q^a=0$
$$\int {\delta I_t\over \delta q^a}\bigg|_{q^a=0} \epsilon^a{}_b \theta^b =
0\eqno(4.2\hbox{b})$$
Since $\theta^a$ is arbitrary, (4.2) shows that $\delta I_t/\delta
q^a\big|_{q^a=0}$ vanishes when the other equations hold.  [This result is true
provided the gravitational variables are dynamical, {\it i.e.\/} they are
included in the set $\chi$ and are present in the total action $I_t$.  The
gravitational action is given in Section~V.]
\goodbreak
\bigskip
\noindent{\bf A.\quad Point Particle}
\medskip
\nobreak
We carry out the above-described procedure for the point-particle action,
whose first-order and Lorentz invariant form is (2.A32).  The Poincar\'e
gauge invariant expression therefore reads
$$I_{\rm particle} = \int d\tau\left\{ p_a \left( {\cal D}_\tau q\right)^a -
{1\over 2} N \left(
p^2 + m^2\right) +  q_A A^A_\mu \dot x^\mu - {1\over 2} q^a
\epsilon_{ab} \dot q^b\right\} \eqno(4.\hbox{A}1\hbox{a}) $$
Here the Poincar\'e coordinate is taken to be a function of $\tau$, so from
(3.39) $\left( {\cal D}_\tau q\right)^a = \left(\dot x^\mu {\cal D}_\mu
q\right)^a = \dot q^a + \epsilon^a{}_b \left( q^b \omega_\mu - e^b_\mu\right)
\dot x^\mu$.  When $q_A A^A$ is expressed in components, (4.A1a) becomes
$$I_{\rm particle} = \int d\tau \left\{\left( p_a + {1\over 2} \epsilon_{ab}
q^b\right)
\left({\cal D}_\tau q\right)^a - {1\over 2} N \left( p^2+m^2\right) - \left(
{\cal A}\omega_\mu + a_\mu - {1\over 2} q_a e^a_\mu\right)\dot x^\mu\right\}
\eqno(4.\hbox{A}1\hbox{b})$$
and reduces to (2.A32) at $q^a=0$.  Performing gauge transformations
according to (3.25), (3.26), (3.36) and (3.40) shows that the action changes
by end-point contributions.
$$I_{\rm particle} \longrightarrow I_{\rm particle}
- \int d\tau{d\over d\tau} \left( {\cal A}\alpha + \beta -
{1\over 2} q_a \theta^a\right) \eqno(4.\hbox{A}2)$$
Dynamical equations are gauge invariant, but the Lagrangian acquires a total
derivative, even when the transformation is global: Owing to the ${1\over 2}
 \dot q_a\epsilon^{ab} q_b$ term, which is like a constant
electromagnetic field in the tangent space, the Lagrangian is not invariant
against translations of $q^a$.

Our formulas can be written in terms of tensors that are covariant under
transformations of the extended group.  The action (4.A1) may be presented as
$$\eqalign{I_{\rm particle} &= \int d\tau
\biggl\{ p_A \left( {\cal D}_\tau q\right)^A -
{1\over 2}
N \left( p_A p^A + m^2\right) +  \left( q_A A^A_\mu \dot x^\mu +
{1\over 2} q^A f_{AB2} \dot q^B\right) \cr
&\quad  - \lambda_1\left( q_A q^A + 2{\cal A}
\right) - \lambda_2 \left( i_Aq^A -1\right)- \lambda_3\left( p_A i^A\right)
\biggr\} \cr} \eqno(4.\hbox{A}3)$$
The $\lambda_i$ are Lagrange multipliers enforcing the special properties of
$q^A$ and $p_A$; {\it viz.\/}  $q_A q^A + 2{\cal A}=0$, $i_A q^A=1$, and
$p_Ai^A=0$.
Non-invariance of the Lagrangian is again seen in the presence of the
non-covariant term: ${1\over 2} q^A f_{AB2} \dot q^B =- {1\over 2}
 q^a \epsilon_{ab} \dot q^b$.  But since the action changes only by
end-point contributions, the equations of motion are gauge
invariant; when  $\lambda_i$ are eliminated they read in terms of $p\equiv
p_A {\cal Q}^A$ and $q\equiv q^A{\cal Q}_A$
$$\eqalignno{{\cal D}_\tau q &= N\Bigl( p +\ll p,q\rr{\cal I}\Bigr)
&(4.\hbox{A}4\hbox{a}) \cr
{\cal D}_\tau p +\ll {\cal D}_\tau p,q\rr {\cal I}
&= \left[{\cal D}_\tau q,q\right] &(4.\hbox{A}4\hbox{b})\cr}$$
Also the gauge current
$$\eqalign{J^\mu\Bigl(x(\tau)|x\Bigr) &= {\cal Q}^A
J^\mu_A\Bigl(x(\tau)|x\Bigr) = - {\cal Q}^A{1\over\sqrt{-g}}\
{\delta I_{\rm particle} \over \delta A^A_\mu(x)}  \cr
&= {1\over\sqrt{-g}} \int d\tau
\Bigl( \left[ p(\tau), q(\tau)\right] - q(\tau)\Bigr)
\dot x^\mu(\tau) \ \delta^2\!\Bigl( x - x(\tau)\Bigr) \cr}\eqno(4.\hbox{A}5)$$
is covariantly conserved.
$${1\over \sqrt{-g}} {\cal D}_\mu \sqrt{-g}\, J^\mu = {1\over\sqrt{-g}}
\partial_\mu \sqrt{-g}\,
J^\mu + \left[ A_\mu, J^\mu\right] = 0\eqno(4.\hbox{A}6)$$
Equation (4.A6) effectively entails,
$${\cal D}_\tau \Bigl([p,q] - q\Bigr) = 0 \eqno(4.\hbox{A}7)$$
which is readily established from (4.A4).  One further equation follows from
varying $x^\mu(\tau)$.
$$\int d^2x \ll F_{\mu\nu}(x), J^\nu\bigl(x(\tau)|x\bigr)\rr=0
\eqno(4.\hbox{A}8)$$
Here $F_{\mu\nu}$ is the gauge curvature (3.35) and (4.A4) as well as
(4.A7) have been used to simplify (4.A8).
\goodbreak
\bigskip
\noindent{\bf B.\quad Fermi Fields}
\medskip
\nobreak
When the steps outlined in the introductory paragraphs
are implemented, the Lorentz gauge invariant
Fermi field action (2.B3) gives rise to the following Poincar\'e gauge
invariant expression, which reduces to (2.B3) in the gauge $q^a=0$.  [Here the
Poincar\'e coordinate is a field: $q^A(x)$.]
$$\eqalign{&I_{\rm Fermi}
 = \int d^2x\, e(q)\Biggl\{ {i\over 2}\bar{\psi}\gamma^\mu(q)
\left(\uprightarrow{\partial}_\mu - i \left( q_A A^A_\mu - {1\over 2} q^a
\epsilon_{ab} \partial_\mu q^b\right)\right)\psi \cr
&\qquad
-{i\over 2}\bar{\psi}\left(\upleftarrow{\partial}_\mu
+ i  \left( q_A A^A_\mu - {1\over 2}
q^a \epsilon_{ab} \partial_\mu q^b\right)\right) \gamma^\mu(q)\psi
- m \bar{\psi} \psi\Biggr\} \cr}\eqno(4.\hbox{B}1)$$
$$\eqalign{
\gamma^\mu(q) &\equiv
{1\over e(q)} \epsilon^{\mu\nu}\left( {\cal D}_\nu q\right)^a\gamma_a\cr
e(q) &\equiv {1\over 2} \epsilon^{\mu\nu} \epsilon_{ab} \left(
{\cal D}_\mu q\right)^a \left({\cal D}_\nu q\right)^b \cr}\eqno(4.\hbox{B}2)$$
where $e(q)$ is a $q$-dependent generalization of $\det e^a_\mu=\sqrt{-g}$,
just as
$-\epsilon^a{}_b \left( {\cal D}_\mu q\right)^b$ generalizes $e^a_\mu$.
Invariance is complete when the Fermi fields transform as
$$\eqalign{\psi &\to e^{{i\over 2} \alpha\gamma_5 - i \left( {\cal A}
\alpha + \beta - {1\over
2} q_a\theta^a \right)}\psi \cr
\bar{\psi} &\to \bar{\psi} e^{-{i\over 2}\alpha\gamma_5 + i \left( {\cal
A}\alpha + \beta - {1\over 2} q_a\theta^a\right)}\cr}\eqno(4.\hbox{B}3)$$
This transformation, which reduces to (2.B5) at $q^a=0$,
is needed to compensate
for the total derivative that arises when $q_A A^A_\mu - {1\over 2} q^a
\epsilon_{ab} \partial_\mu q^b$ is tranformed, see (4.1).

The above formulas are not manifestly covariant.
The route to a covariant formalism is indicated by the remarks in Section~III,
Eqs.~(3.27) -- (3.31): what is needed is a representation of the extended
Poincar\'e group, with ${\cal I}$ realized by a multiple of the identity,
which is necessarily infinite-dimensional. Then the two-component
$\psi$ field and $2\times 2$ Dirac matrices that occur in (4.B1) are obtained
by projection.

The infinite representation of the extended Poincar\'e group that we use
here is given in (3.32) and (3.33).  We also need
an infinite-dimensional representation of the Dirac matrices satisfying
$$\eqalignno{
\left\{ \Gamma_a, \Gamma_b\right\} &= 2h_{ab}\equalsone &(4.\hbox{B}4) \cr
\left[ \Gamma_a, J\right] &= \epsilon_a{}^b \Gamma_b &(4.\hbox{B}5)\cr}$$
A solution is
$$\eqalignno{
\ll n' \left| \Gamma_+ \right|n\rr &= \cases{ 0 & $n$ even \cr\noalign{\vskip
0.2cm}
\sqrt{2}\,\delta_{n',n-1} & $n$ odd \cr} &(4.\hbox{B}6\hbox{a})
\cr\noalign{\vskip 0.2cm}
\ll n'\left| \Gamma_-\right|n\rr &= \cases{ \sqrt{2}\,\delta_{n',n+1} & $n$
even
\cr\noalign{\vskip 0.2cm}0 & $n$ odd\cr} &(4.\hbox{B}6\hbox{b})\cr}$$
We shall also need various anti-commutators.  In the above representation we
find
$$\eqalignno{
\ll n'\left| {1\over 2i} \left\{ \Gamma^a, P_a\right\}\right|n\rr &= M(n)
\delta_{n'n} &(4.\hbox{B}7\hbox{a}) \cr
M(2N) = M(2N+1) &= - i\epsilon \sqrt{2(\nu + 2N+1)}
&(4.\hbox{B}7\hbox{b}) \cr}$$
Also we have
$$\ll n'\left| \left\{ \Gamma_a,J\right\} \right|n\rr = \left( 2j+n'+n\right)
\ll n'\left|\Gamma_a\right|n\rr \eqno(4.\hbox{B}8)$$

We see that $J,M$ and $\Gamma_a$ are $2\times 2$ matrices imbedded along the
diagonal of an infinite matrix.  A set of these $2\times 2$ matrices, given by
fixing adjacent values of
$n$, corresponds to a diagonal Lorentz generator $J$ with eigenvalues
differing by 1, a diagonal $M$ matrix with equal eigenvalues that will become
identified with the fermion mass $m$, and a conventional $2\times 2$
realization
of Dirac matrices.  While it is true that the action of the infinite matrices
representing $P_a$ takes one out of the above-mentioned $2\times 2$
blocks, we shall be
able to disregard this since we shall employ the projection (3.30) to
``neutralize'' the translation generators.

An invariant action is constructed from infinite component fields
$\Phi$, and $\bar{\Phi}$, transforming by
$$\Phi \to U^{-1} \Phi\ \ ,\qquad \bar{\Phi} \to
\bar{\Phi}U\eqno(4.\hbox{B}9)$$
where $U$ is as in (3.5), with generators given by the above infinite
matrices.
$$\eqalign{
I_{\rm Fermi} &= \int d^2x\,e(q)\left\{{i\over 2}\bar{\Phi}
\Gamma^\mu(q)\uprightarrow{{\cal{D}}}_\mu\Phi
- {i\over 2}\bar{\Phi}\upleftarrow{{\cal{D}}}_\mu
\Gamma^\mu(q) \Phi \right\}\cr
\Gamma^\mu (q) &\equiv {1\over e(q)} \epsilon^{\mu\nu}
\left({\cal D}_\nu q\right)^a T^{-1}(q) \Gamma_a T(q) \cr
\uprightarrow{{\cal{D}}}_\mu \Phi &= \partial_\mu \Phi + A_\mu \Phi \cr
\bar{\Phi}\upleftarrow{{\cal{D}}}_\mu &= \partial_\mu \bar{\Phi}-\bar{\Phi}
A_\mu
\cr}\eqno(4.\hbox{B}10)$$
Of course the Lie algebra-valued gauge field and the group element $T(q)$ are
realized in the above infinite-dimensional representation.
It is straightforward to verify that (4.B10)
is invariant; one uses  the fact that when $q$ and the connections in
$\Gamma^\mu(q)$ are transformed,
the result is an adjoint transformation of that quantity.
$$\Gamma^\mu(q)\to U^{-1} \Gamma^\mu (q) U\eqno(4.\hbox{B}11)$$
Note that here we do not use the constrained covariant four-vector
$q^A$; rather we remain with the non-covariant two-component Poincar\'e
coordinate $q^a$.  It enters (4.B10) only through $\Gamma^\mu(q)$, which
however is a covariant object, as is seen from (4.B11).

It is interesting that only a gauged kinetic term is present in the action
(4.B10).   A mass term is not explicitly included; it emerges in the
subsequent reduction.  Following (3.30) we define
$$\Phi = T^{-1}(q)\Psi\ \ ,\qquad \bar{\Phi} = \bar{\Psi}T(q)
\eqno(4.\hbox{B}12)$$
Substituting (4.B12) into (4.B10) and combining the exponentials results in
$$\eqalign{I_{\rm Fermi} &= \int d^2 x \, e(q)\Biggl\{ {i\over 2} \bar{\Psi}
\tilde\Gamma^\mu(q) \left( \uprightarrow{\partial}_\mu - I
\left( q_A A^A_\mu - {1\over 2} q^a \epsilon_{ab}\partial_\mu
q^b\right)\right)\Psi \cr
&- {i\over2}\bar{\Psi} \left(\upleftarrow{\partial}_\mu
+ I \left( q_A A^A_\mu - {1\over 2} q^a
\epsilon_{ab} \partial_\mu q^b\right) \right) \tilde\Gamma^\mu(q)\Psi\cr
&+ {i\over 2}\bar{\Psi} \left\{ \Gamma^a, P_a\right\} \Psi  +{i\over
2}\omega_\mu \bar{\Psi} \left\{ \tilde\Gamma^\mu,J-{\cal A}I\right\}\Psi
\Biggr\}\cr
\tilde\Gamma^\mu(q) &= {1\over
e(q)}\epsilon^{\mu\nu} \left( {\cal D}_\nu q\right)^a\Gamma_a
= T(q) \Gamma^\mu (q) T^{-1} (q)
\cr}\eqno(4.\hbox{B}13)$$
The translational generator here is present only in $\left\{ \Gamma^a,
P_a\right\}$, which according to (4.B7) is diagonal.  Hence the infinite
matrices in (4.B13) are composed of decoupled $2\times2$ blocks, and we can
focus on one of these blocks.

To compare (4.B13) with (4.B1) and also (3.31b) with (4.B2), we look at two
adjacent components of $\Psi$ and call them $\psi\equiv \left( \matrix{
\Psi_{n+1}\cr\Psi_n\cr}\right)$, where for definiteness $n$ is chosen even.
For (3.31b) to agree with (4.B2), we set $I$ equal to $i\equalsone$  {\it
i.e.\/}, $\epsilon^2=i$; also on this two-dimensional subspace, we can define
the restiction of $J$ as
$-{i\over 2}\gamma_5 + i {\cal A}\equalsone$, {\it i.e.\/}
$j = - n - {1\over 2} + i{\cal A}$,
where now $n$ is the fixed number chosen above, and $\gamma_5=i\sigma_3$.
Then with the help of the identities (4.B7) and (4.B8) the action
(4.B13) reproduces (4.B1): the last term in (4.B13) vanishes while the
next-to-last gives the mass in (4.B1) provided we define
$$m = \sqrt{2} \ \sqrt{{\cal A} + i \left( j - \nu -{1\over 2}\right)}
\eqno(4.\hbox{B}14)$$
Therefore the Casimir in this representation is
$$C = \left( 2{\cal A} - m^2\right)\equalsone\ \ .\eqno(4.\hbox{B}15)$$
[Had we not scaled ${\cal B}$ to unity, it would scale the mass as
$m/\sqrt{{\cal B}}$.]

The equations of motion following from (4.B10) by varying $\bar{\Phi}$ and
$\Phi$ are
$$\eqalignno{i\Gamma^\mu(q)\uprightarrow{{\cal {D}}}_\mu
\Phi + {i\over 4} {\epsilon^{\mu\nu}\over e(q)} f_{AB}{}^CF^A_{\mu\nu}
 q^B \ T^{-1}(q) \Gamma_C T(q) \Phi &= 0 &(4.\hbox{B}16\hbox{a})\cr
i\bar{\Phi} \upleftarrow{{\cal{D}}}_\mu
\Gamma^\mu(q) + {i\over 4} {\epsilon^{\mu\nu}\over e(q)}
f_{AB}{}^C F_{\mu\nu}{}^A q^B\ \bar{\Phi} T^{-1}(q) \Gamma_C T(q)  &= 0
&(4.\hbox{B}16\hbox{b}) \cr}$$
In order to present these equations
in an explicitly covariant form, we have introduced four quantities
$\Gamma_A$, coinciding with $\Gamma_a$ for $A=0,1$, unspecified $\Gamma_2$ and
vanishing $\Gamma_3$; the constraint is invariant:
$i^A\Gamma_A=0$.  Thus $T^{-1}(q)\Gamma_A T(q)$ behaves in
the same way as $p_A$ in the particle case, and one further verifies from
(4.B11) the transformation law,
$$T^{-1}(q)\Gamma_a T(q) \to T^{-1} \left( \Lambda^{-1}
(q+\epsilon\theta)\right)
\Gamma_a T\left( \Lambda^{-1}(q+\epsilon\theta)\right) = U^{-1}
T^{-1}(q)\Gamma_b T(q) U \Lambda^b{}_a\eqno(4.\hbox{B}17)$$
which again parallels the behavior of $p_a$, see (3.26).

Varying $q^a$ in (4.B10) leads to
$$i{\epsilon^{\mu\nu}\over e(q)}\bar{\Phi}\upleftarrow{{\cal{D}}}_\mu
T^{-1} (q) \Gamma_a T(q) \uprightarrow{{\cal{D}}}_\nu
 \Phi + {i\over 4}\ {\epsilon^{\mu\nu}\over e(q)} \bar{\Phi}
\left\{ T^{-1}(q) \Gamma_a T(q), F_{\mu\nu}\right\}\Phi =
0\eqno(4.\hbox{B}18)$$
which again is covariant once $\Gamma_a$ is extended to $\Gamma_A$.

The gauge current
$$\eqalign{J^\mu_A &= - {1\over\sqrt{-g}}\ {\delta I_{\rm Fermi}\over \delta
A^A_\mu} \cr
&=-{i\over 2} \ {e(q)\over\sqrt{-g}}
\bar\Phi \left\{\Gamma^\mu(q), Q_A\right\}\Phi
\cr
&+ {i\over 2} f_{AB}{}^C q^B {1\over\sqrt{-g}}\epsilon^{\mu\nu}
\left( \bar\Phi T^{-1}(q)
\Gamma_C T(q) \uprightarrow{{\cal{D}}}_\nu\Phi - \bar\Phi
\upleftarrow{{\cal{D}}}_\nu T^{-1}(q) \Gamma_C T(q) \Phi\right) \cr}
\eqno(4.\hbox{B}19)$$
can be shown to be covariantly conserved
$${1\over \sqrt{-g}}\left( {\cal D}_\mu \sqrt{-g}\,
J^\mu\right)_A = 0 \eqno(4.\hbox{B}20)$$
with the help of equations of motion (4.B16), (4.B18).
\goodbreak
\bigskip
\noindent{\bf V.\quad GRAVITATIONAL ACTION}
\medskip
\nobreak
The gravitational variables that are present in our gauge-theoretical matter
Lagrangians comprise the conventional {\it Zweibein\/} $e^a_\mu$ and spin
connection $\omega_\mu$; also there is the unconventional connection $a_\mu$,
associated with the center in our Poincar\'e algebra.  The gravitational
action provides equations that determine these quantities.  We anticipate that
the {\it Zweibein\/}
equation will encode the torsion free-condition (1.5), thereby
evaluating the spin connection as in (1.4).  Also we anticipate that $a_\mu$
satisfies (2.A7b).  Furthermore, the action must give a determination of the
curvature $R$ in (1.8).  Finally, all these conditions should arise in a gauge
theoretical formalism based on the extended Poincar\'e group.

The above requirements are successfully met in a gravitational action that is
related to the one for ``string inspired, dilaton'' gravity, which reads, in
terms of variables employed originally,$^{3,\,4}$
$$I_{\rm dilaton} = {1\over G}\int d^2x \,\sqrt{-g'}\, e^{-2\varphi} \biggl(
R' - 4\partial_\mu \varphi\partial^\mu \varphi -
\lambda_0\biggr) \eqno(5.1)$$
Here, $G$ is the gravitational coupling strength,  $\lambda_0$ is a
cosmological constant, and temporarily we use a primed metric variable and
scalar curvature to
distinguish them from the unprimed ones, occurring in our formulation.
With the definitions
$$\eqalignno{e^{2\varphi} &= {1\over\eta}&(5.2\hbox{a}) \cr
g'_{\mu\nu} &= {g_{\mu\nu}\over\eta} &(5.2\hbox{b})\cr}$$
the action (5.1) becomes$^1$
$$I_{\rm dilaton} = {1\over G} \int d^2x\sqrt{-g}\, \left( \eta R -
\lambda_0\right) \eqno(5.3)$$
This is similar to the action for the constant curvature de~Sitter model,$^5$
$$I_{\rm de~Sitter} = {1\over G} \int d^2x\,\sqrt{-g}\, \eta\left( R -
\lambda_0\right) \eqno(5.4)$$
but (5.3) differs from (5.4) because $\eta$ does not multiply the
cosmological constant in the dilaton model.  Nevertheless, apart from a
boundary
term, (5.3) may be obtained from (5.4) in a singular limit: shift
$\eta$ in (5.4) by $\lambda'_0/\lambda_0$ and set $\lambda_0$ to zero.$^{11}$
[The
contribution ${\lambda'_0\over\lambda_0}\ {1\over G}\int d^2x\,\sqrt{-g}\,R$
is a boundary term.]

This singular limit has an algebraic parallel.  The de~Sitter algebra, which
leads to the gauge theoretical formulation of (5.4),$^5$ reads
$$\eqalignno{\left[ {\cal P}_a, {\cal P}_b\right]
&= - {1\over 2}\lambda_0 \epsilon_{ab} {\cal J}\ \ ,&(5.5) \cr
\left[{\cal P}_a, {\cal J}\right] &= \epsilon_a{}^b{\cal P}_b &(5.6) \cr}$$
Since ${\cal J}$ is a generator of Abelian rotations, it may be shifted by an
arbitrary multiple of ${\cal I}$.  When ${\cal J}$ is shifted by $-{2\over
\lambda_0}{\cal I}$ and $\lambda_0$ is set to zero, the de~Sitter algebra
(5.5), (5.6) contracts to extended Poincar\'e algebra (3.2), (3.3).

The gauge theoretical formulation of (5.3) uses the quartet of
gauge curvatures $F^A$, constructed from the gravitational
variables and transforming under the adjoint representation, together with a
quartet of Lagrange multiplier fields $\eta_A$, transforming according to the
co-adjoint representation.   Thus the Poincar\'e gauge invariant gravitational
action is$^1$
$$\eqalign{I_{\rm gravity} &= {1\over G}\int \eta_A F^A\cr
&= {1\over G}\int \left(
\eta_a\left( de^a + \epsilon^a{}_b \omega e^b\right)+ \eta_2
d\omega + \eta_3 \left( da + {1\over 2} e^a \epsilon_{ab} e^b\right)\right)
\cr}\eqno(5.7)$$
Setting to zero the variation of $\eta_a$ yields the required torsion
free-condition, while varying $\eta_3$ equates $da$ with the volume two-form.
Finally, variation of $\eta_2$ shows that the curvature vanishes.
[Nevertheless, the model remains non-trivial because the ``physical'' metric
in the dilaton-string context is given by $g_{\mu\nu}/\eta$, see (5.2b).]

The relation between (5.7) and (5.3) is seen in the following manner.  First
we enforce the torsion-free condition and evaluate $\omega$ in (5.7) on (1.4).
Next we vary $a_\mu$, and assume that this quantity {\it does not\/} occur in
the matter Lagrangian; {\it i.e.\/} matter dynamics does not include our
unconventional, dimension-specific force.  The equation that results from
the variation shows that
$d\eta_3$ vanishes, {\it i.e.\/} $\eta_3$ is constant, which we set equal to
$\lambda_0$.  With these steps and the definition $\eta_2\equiv2\eta$ we see
that $I_{\rm gravity}$ becomes $I_{\rm dilaton}$, apart from the boundary term
$\int\lambda_0 da$.

In the more general case, when our
unconventional, dimension-specific force {\it does\/} modify matter dynamics,
the equation for $\eta_3$ coincides with that for $\lambda$ [Eqs.~(2.A16) or
(2.B10)], and $I_{\rm gravity}$ still resembles $I_{\rm dilaton}$, but the
cosmological term is no longer constant.

By varying the Lagrange multiplier $\eta_A$, the gravitational equations can
be presented in gauge invariant form.
$$F^A_{\mu\nu} = 0\eqno(5.8)$$
Varying the connections $A^A_\mu$ gives an equation for the $\eta_A$.
$$\eqalign{{1\over\sqrt{-g}}
\epsilon^{\mu\nu} \left( {\cal D}_\nu \eta\right)_A &= -G J^\mu_A \cr
\left( {\cal D}_\mu\eta\right)_A &= - GJ^{5}_{\mu A}\cr}\eqno(5.9)$$
Here the right side contains the matter gauge current (4.A5) for point
particles or (4.B19) for Fermi fields.

We recall$^2$ that in the absence of matter, the solution for $\eta_A$ produces
the ``black-hole'' metric tensor $g'_{\mu\nu} =2 g_{\mu\nu}/\eta_2$, while the
invariant $\eta_A\eta^A$ determines the product of the ``black-hole'' mass
with the cosmological constant $\lambda_0$.

The most direct solution to (5.8)
$$A^A_\mu=0\eqno(5.10)$$
simplifies drastically the matter equations, while (5.9) reduces to
$$\partial_\mu \eta^0_A = -
GJ^5_{\mu A}\bigg|_{A^A_\mu=0}\eqno(5.11)$$
Of course, this solution is highly singular from the geometrical point of
view: since $A^A_\mu$ vanishes, so do the {\it Zweibein\/}, spin-connection
and $a_\mu$; no geometry can be described.  However, geometric content is
easily regained by performing a gauge transformation on (5.10),
$$A_\mu \equiv A^A_\mu Q_A\longrightarrow U^{-1} \partial_\mu U \eqno(5.12)$$
where $U$ is arbitrary, but chosen so that geometrical quantities are
non-singular.  After the gauge transformation  $H^0\equiv \eta^0_A Q^A$ becomes
replaced by $H = U^{-1} H^0U$, but the invariant content of a solution,
coded {\it e.g.\/} in $\eta_A \eta^A$, remains unchanged.  In the absence of
matter (5.11) is solved by four constants comprising $\eta^0_A$, which
correspond to the cosmological constant, as well as to
the black hole location in space-time and its mass.
\goodbreak
\bigskip
\noindent{\bf VI.\quad CONCLUSIONS}
\medskip
In this work we have given details not included in our two recent
Letters$^{1,\,2}$ about a gauge theory of lineal gravity built on the
centrally extended Poincar\'e group.  We have
also considered the coupling of matter points and fields, whose gauge
invariance is seen explicitly in our manifestly covariant tensor
calculus.  Within this formalism, additional non-standard and
dimension-specific matter gravity interactions are elegantly incorporated.
This is accomplished by using the connection $A^A_\mu$ and the Poincar\'e
four-vector $q^A$, whose presence in the action in the combination (4.1) gives
rise to the additional forces, which are also seen to be correlated with the
extension of the Poincar\'e group.  The strength ${\cal A}$ of
the curvature-dependent force is determined by the length of $q^A$ and can
take any value.  Since the quantity (4.1) may be deleted from the action
without upsetting gauge invariance, it is
recognized that the additional forces need not be included in the matter
Lagrangian, even when the extended group is used for the gravity sector.
Finally, we call attention to the fact that the cosmological constant in
gauged gravity arises dynamically, and is not a parameter in the Lagrangian.
Similar mechanisms for generating a cosmological constant have been previously
posited in the physical four-dimensional theory,$^{12}$ but for us this is
dictated by group structure.  Moreover, our additional interactions result in
space-time variation of the cosmological term.
\vfill
\eject
\centerline{\bf REFERENCES}
\medskip
\item{1.}D.~Cangemi and R.~Jackiw, {\it Phys. Rev. Lett.\/} {\bf 69}, 233
(1992).
\medskip
\item{2.}D.~Cangemi and R.~Jackiw, {\it
Phys. Lett. B\/} {\bf 299}, 24 (1993).
\medskip
\item{3.}C.~Callan, S.~Giddings, J.~Harvey and A.~Strominger, {\it Phys. Rev.
D\/} {\bf 45}, 1005 (1992).
\medskip
\item{4.}H.~Verlinde, in {\it Sixth Marcel Grossmann Meeting on General
Relativity\/}, M.~Sato, ed. (World Scientific, Singapore, 1992).
\medskip
\item{5.}C.~Teitelboim, {\it Phys. Lett.\/} {\bf 126B}, 41 (1983); in {\it
Quantum Theory of Gravity\/}, S.~Christensen, ed. (Adam Hilger, Bristol,
1984); R.~Jackiw in {\it ibid.\/}; {\it Nucl. Phys.\/} {\bf B252}, 343 (1985).
 T.~Fukiyama and K.~Kamimura, {\it Phys. Lett.\/} {\bf 160B}, 259 (1985);
K.~Isler and C.~Trugenberger, {\it Phys. Rev. Lett.\/} {\bf 63}, 834 (1989);
A.~Chamseddine and D.~Wyler, {\it Phys. Lett.\/} {\bf B228}, 75 (1989).
\medskip
\item{6.}A review of the gauged Poincar\'e group in gravity theory is by
T.~Kibble and K.~Stelle, in {\it Progress in Quantum Field Theory\/}, H.~Ezawa
and S.~Kamefuchi, eds. (North-Holland, Amsterdam, 1986).
\medskip
\item{7.}Alternatively, following G. Grignani and G. Nardelli, {\it Nucl.
Phys. B\/} (in press),  one may introduce
additional fields --- the ``Poincar\'e coordinates''
in terms of which cosmological
constant effects  can be formulated gauge invariantly. We use these
Poincar\'e coordinates in Section~IV
to describe matter interactions in a gauge invariant
fashion, but we do not introduce them in the gravity sector.
\medskip
\item{8.}H.~Leutwyler, {\it Phys. Lett.\/} {\bf 153B}, 65 (1985); (E) {\bf
155B}, 469 (1985).
\medskip
\item{9.}Various properties of $Os(1)$ are discussed by W.~Miller, {\it
Commun. Pure and Appl. Math.\/} {\bf 18}, 679 (1965), {\bf 19}, 125 (1965);
R.~Streater, {\it Commun. Math. Phys.\/} {\bf 4}, 217 (1967).
\medskip
\item{10.}This observation
is based on related remarks by Grignani and Nardelli;  Ref.~[7].
\medskip
\item{11.}A.~Ach\'ucarro, Tufts University preprint (July 1992).
\medskip
\item{12.}A.~Aurilia, H.~Nicolai and P.~Townsend, {\it Nucl. Phys.\/} {\bf
B176}, 509 (1980); S.~Hawking, in {\it Shelter Island II\/}, R.~Jackiw,
N.~Khuri, S.~Weinberg and E.~Witten, eds. (The MIT Press, Cambridge, MA,
1985);   M.~Henneaux and C.~Teitelboim, {\it Phys. Lett.\/} {\bf 143B}, 415
(1984), {\bf B222}, 195 (1989).
\par
\vfill
\end